\documentclass[fleqn,usenatbib]{mnras}

\usepackage[T1]{fontenc}
\usepackage{ae,aecompl}

\usepackage{graphicx}  
\usepackage{amsmath}  

\newcommand{\eht}{\overline}    
\newcommand{\fht}{\widetilde}

\def\ff#1{#1''}

\def\erho{\eht{\rho}}

\def\av#1{\overline{#1}}
\def\fav#1{\widetilde{#1}}

\def\ie{{\it i.e. }}
\def\eg{{\it e.g. }}
\def\ms{\, {M_{\odot}}}

\def\la{\hbox{\raise.5ex\hbox{$<$} 
    \kern-1.1em\lower.5ex\hbox{$\sim$}}} 
\def\ga{\hbox{\raise.5ex\hbox{$>$} 
    \kern-1.1em\lower.5ex\hbox{$\sim$}}}

\newcommand{\Msun}{\mbox{M$_\odot$\,}}         
\newcommand{\cms}{\mbox{\ cm s${}^{-1}$}}    

\DeclareMathAlphabet{\mathpzc}{OT1}{pzc}{m}{it}

\title[Turbulent Mixing and Nuclear Burning in Stellar Interiors]{Turbulent Mixing and Nuclear Burning in Stellar Interiors}

\author[Miroslav Moc\'ak]{
Miroslav Moc\'ak,$^{1}$\thanks{E-mail:miroslav.mocak@gmail.com}
Casey Meakin,$^{2,3}$\thanks{E-mail:casey.meakin@gmail.com}
Simon Campbell$^{4,5}$\thanks{E-mail:simon.campbell@monash.edu}
and W. David Arnett $^{3}$\thanks{E-mail:wdarnett@gmail.com}
\\
$^{1}$Theoretical Division, LANL, Los Alamos NM 87545, USA\\
$^{2}$Karagozian \& Case, Inc., Glendale, CA 91203\\
$^{3}$Steward Observatory, University of Arizona, Tucson, AZ 85721\\
$^{4}$Monash Centre for Astrophysics, School of Physics and Astronomy,
Monash University, Clayton, Australia 3800\\
$^{5}$Max-Planck-Institut f{\"u}r Astrophysik (MPA), D-85748 Garching, Germany
}

\date{Submitted to MNRAS}

\pubyear{2018}

\begin{document}
\label{firstpage}
\pagerange{\pageref{firstpage}--\pageref{lastpage}}
\maketitle

\begin{abstract}
The turbulent burning of nuclei is a common phenomenon in the
  evolution of stars.  Here we examine a challenging case: the merging of
  the neon and oxygen burning shells in a 23\,\Msun star.  A previously
  unknown quasi-steady state is established by the interplay between
  mixing, turbulent transport, and nuclear burning.  The resulting stellar
  structure has two burning shells {\em within a single convection
    zone}. We find that the new neon burning layer covers an extended
  region of the convection zone, with the burning peak occurring
  substantially below where the Damk\"ohler number first becomes equal to
  unity. These characteristics differ from those predicted by 1D stellar
  evolution models of similar ingestion events. We develop the mean-field
  turbulence equations that govern compositional evolution, and use them to
  interpret our data set. An important byproduct is a means to quantify
  sub-grid-scale effects intrinsic to the numerical hydrodynamic
  scheme. For implicit large eddy simulations, the analysis method is
  particularly powerful because it can reveal where and how simulated flows
  are modified by resolution, and provide straightforward physical
  interpretations of the effects of dissipation or induced
  transport. Focusing on the mean-field composition variance equations for
  our analysis, we recover a Kolmogorov rate of turbulent dissipation
  without it being imposed, in agreement with previous results which used
  the turbulent kinetic energy equation.
\end{abstract}




\section{Introduction}
\label{s-intro}

It is now feasible to simulate stellar convection in three-dimensions
(3D), with realistic microphysics, multiple species of nuclei, and
sufficient resolution in space and time to represent turbulent flow
\citep{MeakinArnett2007,Mocak2011,Woodward2013}.  Historical work on
stellar convection \citep{bohmvitense1958} and 3D simulations of stellar
atmospheres \citep{steinnordlund1998} have generally focused on flows
having uniform composition, a case which is usually appropriate for the
outer layers of stars. 

By contrast convection in stellar interiors is generally characterized by
nuclear burning and nonuniform composition.  Here we examine the
interaction between turbulent convection, thermonuclear burning, and entrainment at
 boundaries. Simulations of convective
shells, driven by nuclear burning, show entrainment of material from
surrounding stable layers. Erosion at boundaries
introduces inhomogeneities in composition, entropy, and buoyancy into the
convective flow. This can be viewed as a multi-stage process of
entrainment, transport, dispersion (or stirring) induced from the largest to
smallest eddies, as well as diffusion, spanning the full spectrum of
space-time scales of the flow \citep{Dimotakis2005}. The feedback of such
mixing on nuclear burning, convection, and its impact
on the evolution of the star remains largely unexplored:
an ad hoc diffusion operator is almost universally used in stellar evolution.
This paper begins to analyze these stages.

We focus on the oxygen burning shell in a
massive supernova progenitor. Oxygen burning and neon burning occur at
sufficiently similar temperatures that these burning shells may interact
\citep{wdaNe74,wdaO74}. Interaction was indeed found in the 3D simulations of
\citet{meakinphd} and \citet{MeakinArnett2007}, but was not analyzed in detail 
there. Here we present a detailed account of the 
compositional mixing and modified nuclear burning, which occurs as the
convective oxygen-burning shell merges with the (initially) stable overlying neon
shell.

We use 3D numerical hydrodynamics in the implicit large eddy simulation
framework ILES \citep{ILES}, which means we solve the Euler equations with
a non-oscillatory finite volume numerical fluid solver. In the current study we use the PPM method \citep{ColellaWoodward1984}.
The dissipation in such solvers comes from solving the Riemann shock problem over
each zone $\Delta_s$, giving a dissipation rate\footnote{The change in specific kinetic energy over the shock traversal  time gives the dissipation rate. PPM approximates sub-grid structure as piece-wise parabolic, smoothing higher-order terms and decreasing information (complexity). Variances in velocity and in scalar variables dissipate/diffuse at this same rate at the sub-grid scale.} of $\sim v_s^3/\Delta_s$ as a shock of speed $v_s$ traverses a zone. In a turbulent cascade the mean damping is ${\cal D} \sim v_s^3/\Delta_s \sim v^3/\ell,$ which is determined by the rms velocity $v$ and dimension $\ell$  of the turbulent region.  
Use of such solvers introduces an implicit sub-grid model which corresponds to a Kolmogorov  turbulent cascade, freeing computational resources to capture the large scales relevant to astrophysics. 
See \cite{ILES} for references and a more rigorous discussion.

This is in contrast to the direct numerical simulation approach DNS
\citep{Pope2000}, which solves the Navier-Stokes difference equations on
the grid, all the way down to the dissipation scale. In this approach most
of the computational effort is spent on these small scales, which are
buried in the turbulent cascade.  \cite{sytine2000} showed that both
methods converge to the same result, but that ILES is more efficient for
highly turbulent flows. We have confirmed that our simulations extend from
the integral (large) scale down into the inertial range of the turbulent
cascade (e.g. \citealt{Cristini2017}).

In this paper, we extend our analysis by using an approach inspired by Reynolds-averaged Navier-Stokes (RANS) methods \citep{VialletMeakin2013,mmva14,amvclm15}. Our approach differs from traditional RANS
\citep{Besnard1992,Chassaing2010} in two fundamental ways: (1) the fluctuations are taken from our simulations, and hence are dynamically constrained, and (2) we solve the ILES Euler equations.
A more accurate acronym than ``RANS'' (which we have used previously) is needed for clarity; we choose ``Reynolds averaged ILES'' (RA-ILES), to distinguish our approach.
{\em Unlike unconstrained RANS analysis, our RA-ILES equations are complete, exact to the accuracy of our grid,  and require little added computational cost.} There is no closure issue\footnote{There is no explicit Navier-Stokes viscosity term to generate higher order moments in the conventional way \citep{tritton1988};  the implicit turbulent cascade gives closure.}. 

Previous papers \citep{MeakinArnett2007,VialletMeakin2013,amvclm15,Cristini2017}
have focused on the turbulent kinetic energy equation (TKE);
here our analysis shifts to mean-field transport equations for the density of $^{16}$O and
$^{20}$Ne, and their turbulent fluxes and variances, complemented by
analysis of relevant timescales and nuclear burning processes.
These transport equations are the ones which deal with changes in composition variance: i.e., mixing.

The paper is organized as follows: 
In \S\ref{sect:simulation-model} we describe the initial conditions and the 3D
stellar model that we investigate in this paper. In
\S\ref{sect:rans-definitions} we develop the RA-ILES equations used for our
analysis. In \S\ref{sect:timescales} we define several timescales which we
use to characterize physical processes operating in the simulated flow. In
\S\ref{sect:results} we present the results from our RA-ILES analysis of oxygen
and neon entrainment, transport, dissipation, and burning. We use the
composition related mean-field equations and provide a systematic
description of each term in the budget equations with an emphasis placed
on physical interpretation. We then look at effects of resolution on 
our results in \S\ref{sect:resolution}. Finally, we conclude with a
summary and discussion in \S\ref{sect:summary}.

\begin{figure*}
\includegraphics[width=0.49\hsize]{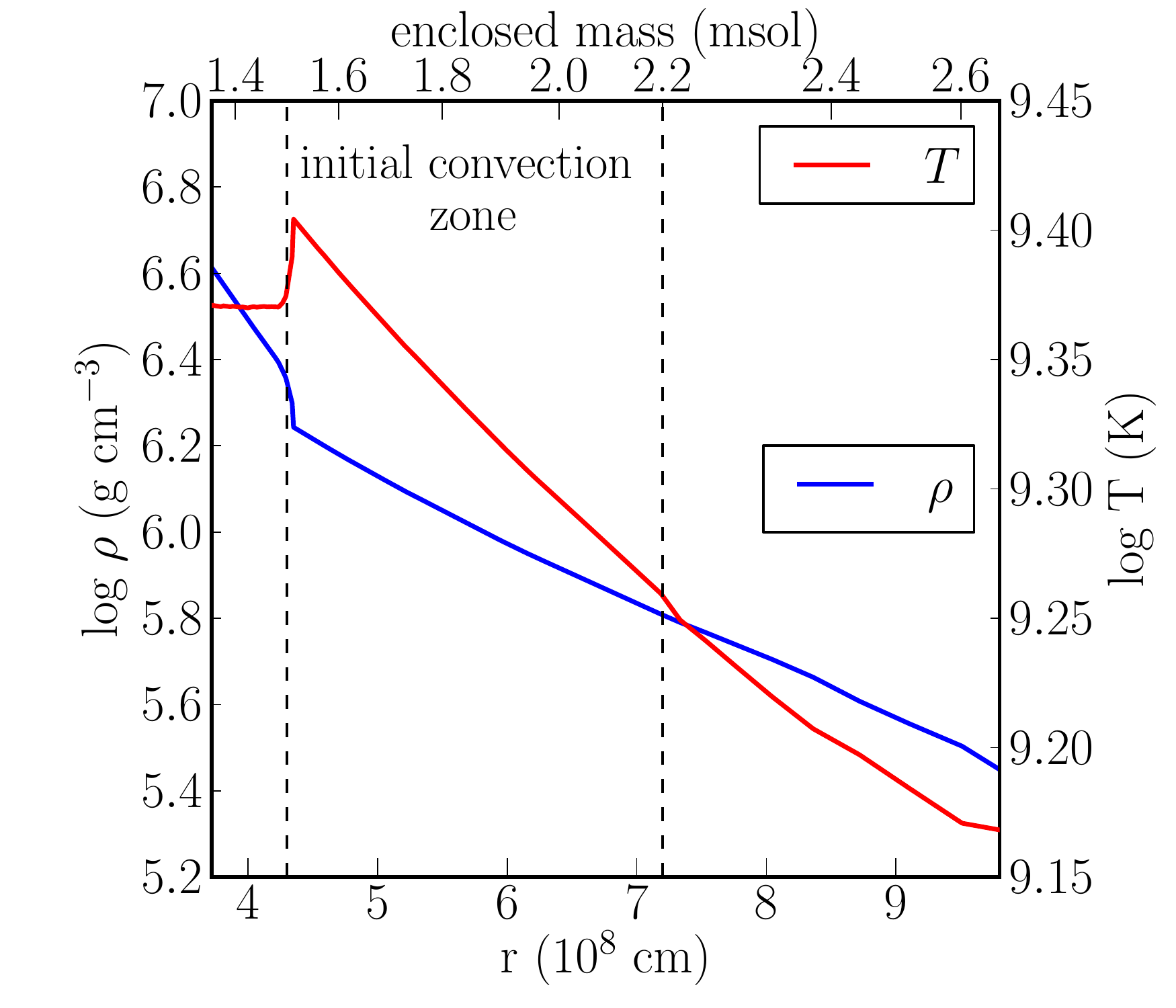}
\includegraphics[width=0.49\hsize]{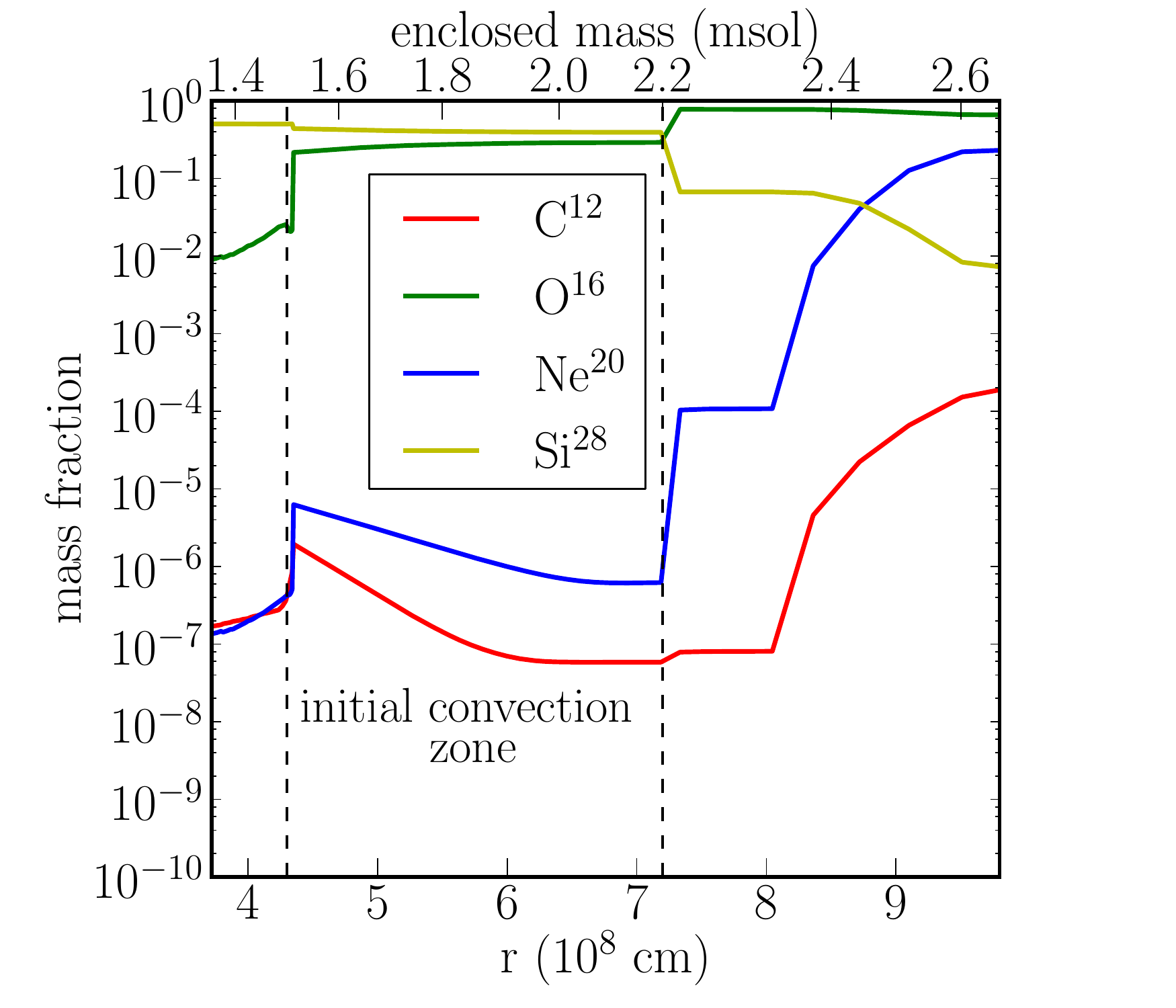}
\includegraphics[width=0.49\hsize]{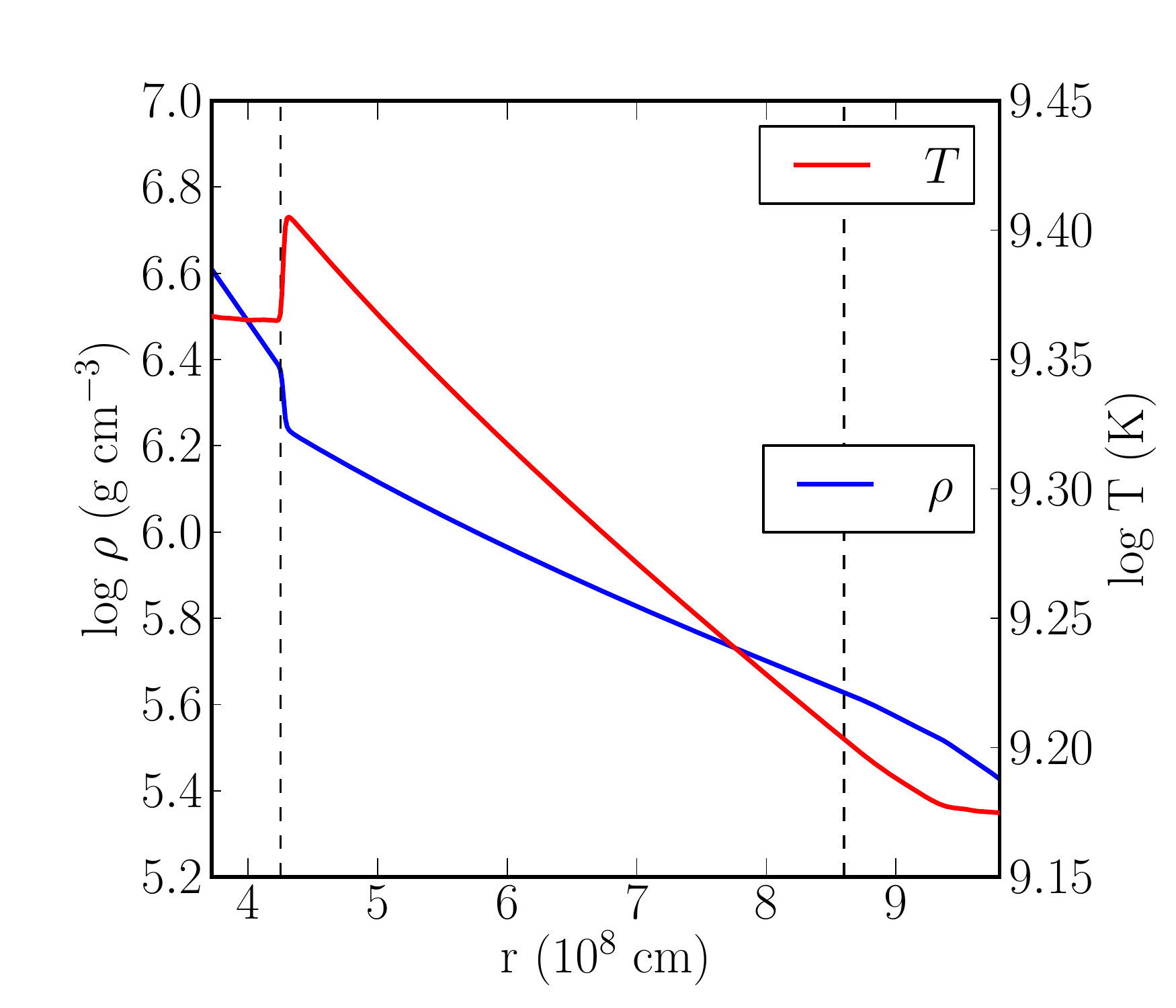}
\includegraphics[width=0.49\hsize]{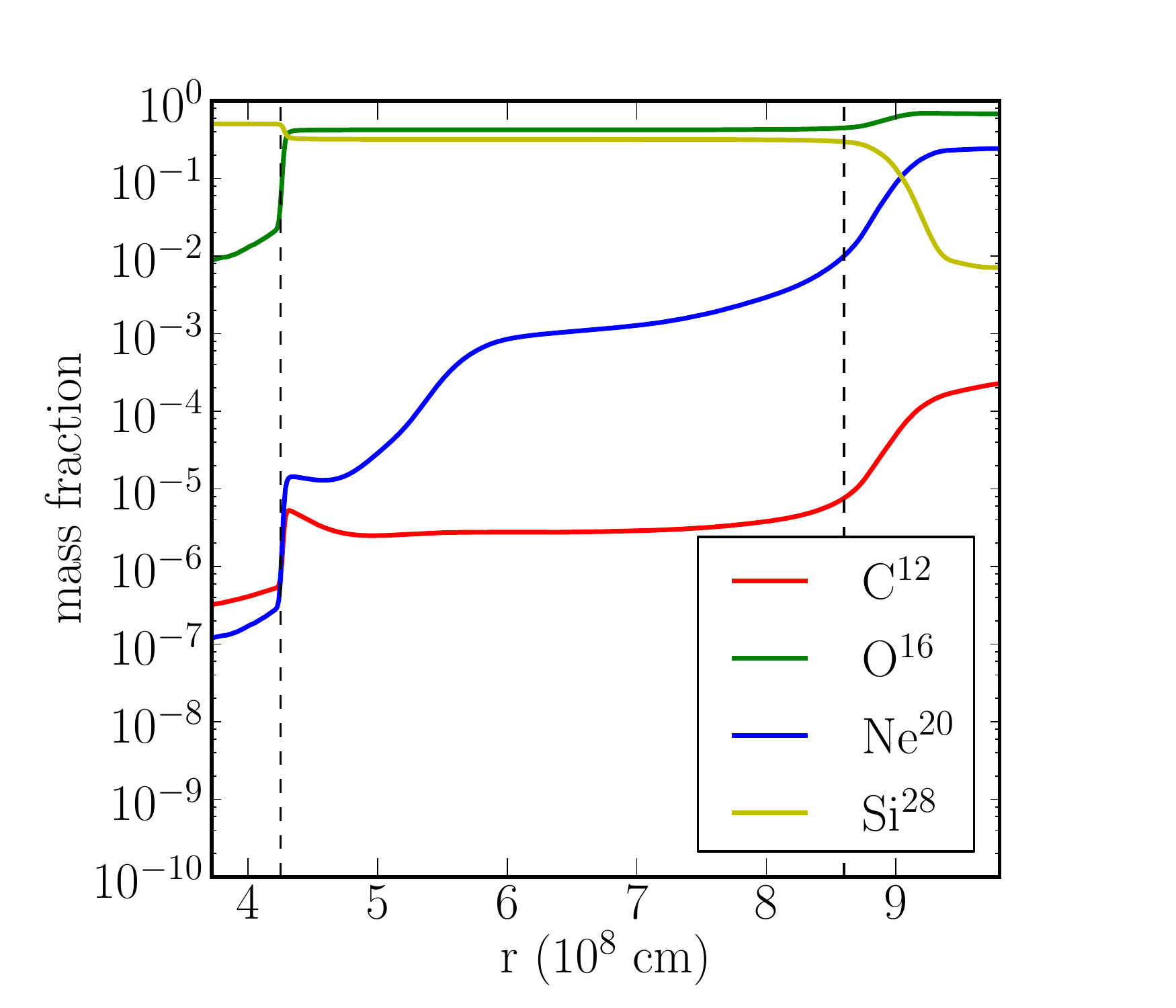}
\caption{{\texttt Top row:} ({\em left}) Initial 1D background temperature $T$ and density
  $\rho$; and ({\em right}) composition profiles (mass fraction).
  {\texttt Bottom row:} Radial profiles after 300 seconds of evolution at which point the model has  obtained a quasi-steady character.
  Only the most energetically important nuclear species are
  shown. Vertical dashed lines mark boundaries of convection.}
\label{fig:initial-model}
\end{figure*}

\section{Initial Model and Simulation method}
\label{sect:simulation-model}


The initial 1D model for our 3D simulation is a 23$\ms$ supernova
progenitor.  It was evolved with the 1D TYCHO stellar evolution code
\citep{YoungArnett2005,ArnettMeakin2010} using mixing-length  theory (MLT) 
to a point just following core
oxygen burning where oxygen, neon, carbon, helium, and hydrogen are burning
in concentric shells outside a degenerate core of silicon and sulfur. The
structure of the 1D initial model is shown in the panels of the top row of 
Figure~\ref{fig:initial-model}. It has a single convective oxygen shell enclosed by two stable
  layers. The initial convective region is driven primarily by nuclear burning of
  oxygen and extends from its base around 4.3$\times
  10^8$ cm up to a radius around 7.2$\times 10^8$ cm, where the Ledoux criterion indicates
  a stable boundary. The stable layers
  below and above are composed primarily of silicon and oxygen; in the top layer the
  dominant nuclear burning is that of
  neon, and nonconvective. Such stellar structures are quite common and can be found, e.g., in cores of low-mass red giants during the core
  helium flash, or during core carbon flashes of ``super-AGB'' stars
  \citep{Mocak2012}.

\begin{table}
\begin{centering}
\begin{tabular}{lc}
\hline
Grid dimensions ($\Delta r,\Delta\theta,\Delta\phi$) &  $6\times 10^{8} \mathrm{cm} \times 27.5^{\circ} \times 27.5^{\circ}$ \\
Grid zoning  & $400 \times 100 \times 100$ \\
$\Delta t_\mathrm{av}$ (s) & 300 s\\
$v_\mathrm{rms}$  & $\sim 1.4 \times 10^7$ cm/s\\
$\tau_\mathrm{conv}$  &  $\sim 65$ s\\
\hline
\end{tabular}
\caption{\label{tab:ob-model} 3D oxygen burning simulation properties
  (model ob.3d.B). $\Delta t_\mathrm{av}$ is the averaging timescale
  for the mean field analysis; $v_\mathrm{rms} = \sqrt{2 K_{tot}/M}$ is
  the approximate global rms velocity (where $K_{tot}$ is the total turbulent
  kinetic energy in the convection zone and $M$ mass contained in the
  convection zone), and $\tau_\mathrm{conv} = 2 l_{cvz}/v_\mathrm{rms}$ is
  the convective turnover timescale (where $l_{cvz}$ is size of the
  convection zone, $\sim 4.3\times 10^8$ cm). All values were obtained at the
  central simulation time 1060 seconds around which we perform all subsequent
  time-averaging.}
\end{centering}
\end{table}

\begin{figure*}
\includegraphics[width=1.0\hsize]{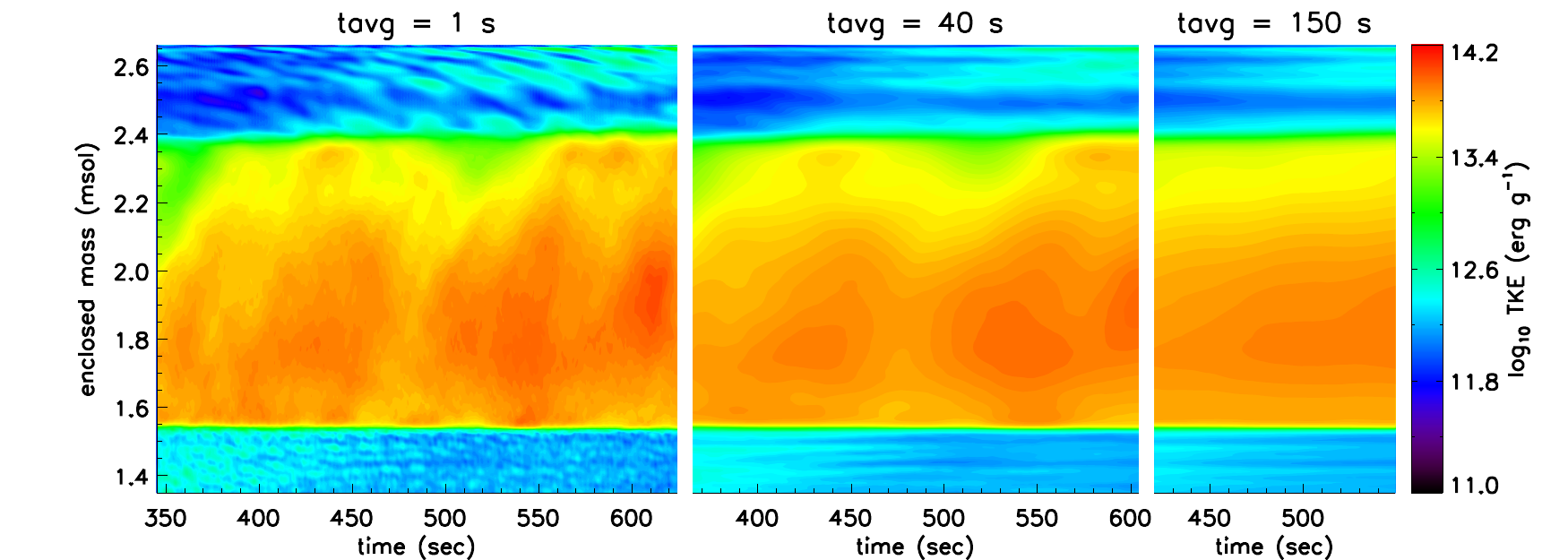}
\caption{Comparison of different averaging windows for turbulent kinetic
  energy and its resulting mean-field. The purpose of averaging is to
  separate short time variability, which tends to cancel itself, from more
  robust long time changes. The left panel illustrates the effect of a
  short averaging window of 1 second; the turbulent kinetic energy is
  modulated by convective bursts, and the boundary layers support g-mode
  waves. The center panel, with an averaging time of 40 seconds, shows
  considerably more smoothing while the right panel, with an averaging time
  of 150 seconds, shows no convective bursts, just steady
  convection.} \label{fig:averaging}
\end{figure*}

Reactive-hydrodynamic evolution in 3D was computed with the PROMPI
code, a version of the legacy PROMETHEUS code \citep{fryxell1991} adapted
to parallel computing via the Message Passing Interface (MPI). This code is
an Eulerian implementation of the piecewise parabolic method (PPM) of
\citet{ColellaWoodward1984} updated with a Riemann solver for real
  gases according to \citet{colella1985}.  Additional added physics include self gravity in the spherically symmetric approximation, a
realistic equation of state to handle the semi-degenerate stellar plasma
\citep{TimmesSwesty2000}, and a general nuclear reaction network. For the
current simulation we use a 25-isotope network that includes neutrons,
protons, $^4$He, $^{12}$C, $^{16}$O, $^{20}$Ne, $^{23}$Na, $^{24}$Mg,
$^{28}$Si, $^{31}$P, $^{32}$S, $^{34}$S, $^{35}$Cl, $^{36}$Ar, $^{38}$Ar,
$^{39}$K, $^{40}$Ca, $^{42}$Ca, $^{44}$Ti, $^{46}$Ti, $^{48}$Cr, $^{50}$Cr,
$^{52}$Fe, $^{54}$Fe, $^{56}$Ni.  All of the important strong and weak
reactions are included. It is conceptually convenient to
  decompose the network results into separate processes. In Appendix \S\ref{nuclearapprox} we give simpler approximations which are used to identify the different stages of burning. 
  
In the simulation neither Si nor C burning are ever
the primary nuclear process. The dominant energy release is due to burning of $^{16}$O
and $^{20}$Ne. The dominance of neon burning is the result of
entrainment as it mixes $^{20}$Ne into the deeper, hotter layers of the oxygen burning
shell. Oxygen burning is invigorated by mixing in new $^{16}$O fuel from the previously non-convective region \citep{MeakinArnett2007}.

We focus on the region encompassing the oxygen and neon burning shells and their interactions with each other and with the adjacent stably stratified layers. The computational grid is defined in a spherical coordinate system with periodic boundary conditions in the angular directions and reflecting (non-transmitting) boundaries in the radial directions.  The opening angle of the grid is $27.5^{\circ}\times 27.5^{\circ}$ in $\theta$ and $\phi$. A short summary of the simulation properties, including zoning,  is presented in Table~\ref{tab:ob-model}. 


Due to its lower computational cost, this medium resolution simulation could be extended over a longer time span than higher resolution simulations, and attain a new and reliable quasi-steady state.

\section{Mean-Field Evolution Equations}
\label{sect:rans-definitions}

In this section we describe and develop the Reynolds-Averaged ILES
  (RA-ILES) analysis for composition \ie mean-field equations describing the
  evolution of mean fields related to composition on spherically averaged
  shells.  The evolution equations for mean composition
  (\S\ref{sect:x-evolution}), turbulent composition flux
  (\S\ref{sect:flux-evolution}), and composition variance
  (\S\ref{sect:variance-evolution}) are then considered in turn.

\subsection{Reynolds-Averaged ILES Analysis}

The RA-ILES equations derived here are exact, and do not employ
  approximations, at least to the extent that the continuum approximation
  is appropriate and flow features are resolved. This contrasts with
  methods using closure relationships and truncations of the RANS equations
  to construct approximate models of turbulence. The RANS equations, when
  closed by 3D numerical simulations to become ``RA-ILES'' equations, are able
  to represent the full range of hydrodynamical behavior present in a
  stellar interior. 

As shown in \cite{hinze75}, \cite{Besnard1992}, and \cite{Chassaing2010},
the RANS framework provides a rational approach to interpreting complex 3D
fluid dynamical data. The extension of the methodology to ``RA-ILES'' (see
\S\ref{s-intro}) for stellar interiors is discussed in
\cite{VialletMeakin2013}, \cite{mmva14}, and \cite{amvclm15}.

\subsection{The Averaging and Decomposition Procedure}
\label{sect:averaging}

In this section we define the averaging rules needed to obtain 1D mean
 and fluctuation fields from 3D flow data. {\em This decomposition makes
  precise the relationship between the 1D mean fields evolved by a stellar
  evolution code and 3D hydrodynamic simulations}
\citep{VialletMeakin2013}. Two types of averaging are used: 
in time, and in space. 
In practice, both types of averaging are combined and
contribute to a meaningful 
average of the flow within a
spherical layer by virtue of the ergodic hypothesis\footnote{An
    average of a physical quantity over a statistical ensemble is equivalent to
    its average over time.} and spherical symmetry\footnote{If spherical
  symmetry were broken, as it would be in the presence of a magnetic field
  or a mean stellar rotation field, spherical averaging would not be
  appropriate. In the case of a mild rotation field, for example, mean
  fields could not be restricted to less than two dimensions and would be
  represented most naturally in a meridional plane.}.

 The 
 average of a quantity $q$ on a spherical shell at radius $r$ (\ie a mean field) is defined by

\begin{equation}
\label{eq:eht}
\eht{q}(r,t_c) = \frac{1}{T\Delta\Omega}\int_{t_c- T/2}^{t_c+T/2} \int_{\Delta\Omega} q(r,\theta,\phi,t')~d\Omega~dt'
\end{equation}

\noindent where $d \Omega = \sin \theta d \theta d \phi$ is the solid angle in spherical coordinates, $T$ is the averaging time period, $t$ is time and $\Delta\Omega$ is the  solid angle being averaged over. The time coordinate $t_c$ represents the center of the time-averaging window used (central time).

Fluctuations, which retain the full time and space dependence as the original, self-consistent 3D flow field,  are defined according to the decomposition $q(\vec{x},t) = \eht{q}(r,t) + q'(\vec{x},t)$, noting that $\eht{q'}(r,t) = 0$ by construction. Similarly, a Favre (or density weighted) average is given by

\begin{equation}
\fht{q} = \eht{\rho q} / \eht{\rho}
\end{equation}

\noindent which defines a complimentary decomposition of the flow according to $q = \fht{q} + \ff{q}$ where $\ff{q}(r,\theta,\phi,t)$ is referred to as the Favrian fluctuation and its mean is zero when Favre averaged: $\fht{\ff{q}}(r,t) = 0$. A more complete elaboration on the algebra of these averaging procedures can be found in \citet{Chassaing2010}. 

The mean fields presented in this paper were calculated by post-processing snapshots (that were written to disk every 0.5 s of simulated time, for a period from 888 s up to 1211 s), by following these steps:


\begin{enumerate}
\item Calculate Reynolds fluctuations for each snapshot, at time $t$, from the raw 3D simulation data, as defined above:

\begin{equation}
q'_{inst}(r,\theta,\phi,t) = q(r,\theta,\phi,t) - \langle q \rangle (r,t)
\end{equation} 

\noindent where the dependence on space and time variables are shown for each term. $\langle q \rangle (r,t) = 1/\Delta\Omega \int_{\Delta\Omega} q(r,\theta,\phi,t)~d\Omega$ gives spatial average of $q$ for a given time $t$ and radius $r$.

\item Calculate needed $products$ of any required thermodynamic quantities $q_1, q_2, q_3$, etc. (for example $q_1' q_3$ or $q_1'' q_2'' q_3''$) from 3D fields. Favrian fluctuation can be obtained from

\begin{align}
q''_{inst} = q'_{inst} - \frac{\langle \rho q'_{inst} \rangle}{\langle \rho \rangle}.
\end{align}

\item space and time average the products calculated in the prior step around central time $t_c$ as defined in Eq.~\ref{eq:eht} above.

\end{enumerate}

We find that the turbulent flow can be very well sampled when using an
averaging time window of width $T$ around two convective turnover timescales
\citep{mmva14}. Figure \ref{fig:averaging} shows an example of the
averaging window effect on our flow for the mean specific turbulent kinetic energy
(TKE) defined as $\fht{k} = \frac{1}{2} (\fht{u_r^{''} u_r^{''}}+\fht{u_\theta^{''}
  u_\theta^{''}}+\fht{u_\phi^{''} u_\phi^{''}})$ where $u_i$ are components
of the velocity field. Small timescale features are absorbed into the
mean-field when the averaging window exceeds around 150 seconds
($\sim$2 turnover times). 
In order to present a robust
  statistical analysis we use an averaging window of 300 seconds
  throughout this paper, which is $\sim$4 turnover times.

\subsection{Evolution Equation for Mean Composition}
\label{sect:x-evolution}

The instantaneous evolution equation for mass fraction of element $i$ in spherical geometry
is (\cite{wda96}, Eq~4.97),
\begin{equation}
\partial_{t} \big(\rho X_{i}\big) =  -\nabla \cdot ( \rho {\bf u} X_i ) + \rho \dot{X}_{i}^{\rm nuc}. 
\label{eq:cont1}
\end{equation}
Applying our decomposition and averaging procedure (Sect.\ref{sect:averaging}), 
we obtain the following 1D transport equation
\begin{equation}
\erho\fav{D}_t \fav{X}_i =  -\nabla_r f_i + \av{\rho}\fav{\dot{X}}_i^{\rm nuc} + {\mathcal N_{i}}
\label{eq:transport-comp}
\end{equation}
\noindent where $X_i$ is mass fraction of chemical element $i$, $\rho$ is density, ${\bf u} = [u_r,u_\theta, u_\phi]$ is the velocity vector, $\nabla$ is the divergence operator, $\dot{X}_i^{\rm nuc}$ is the rate of nuclear burning of $i$, and $f_i = \eht{\rho}\fht{X''_i u''_r}$ is the turbulent flux of element $i$. $\nabla_r (.) = (1/r^2)\partial_r r^2 (.)$ is the radial divergence operator and the $\fht{D}_t$ is the mean-flow Lagrangian derivative $\fht{D}_t {(.)} = \partial_t (.) + \fht{u}_n \partial_n (.)$. 

The mean-field transport equation (Eq.~\ref{eq:transport-comp}) states
that the temporal change of mass fraction of an element $i$ in the
Lagrangian frame of reference, $\erho\fav{D}_t \fav{X}_i$, is caused by either
a spatial redistribution by the the turbulent flux, $-\nabla_r f_i$; or
by nuclear burning, $\av{\rho}\fav{\dot{X}}_i^{\rm nuc}$. 

We define the
numerical residual in these equations by ${\mathcal N_{i}}$, which
    represents the implicit action of the numerical simulation algorithm. These terms 
    are discussed in more detail in the following sections and again later in \S\ref{sect:results}.

\subsection{Evolution Equation for Turbulent Composition Flux}
\label{sect:flux-evolution}

The transport equation for the turbulent flux of an arbitrary chemical element can be obtained by using the following general formula for second-order moments \citep{mmva14},

\begin{align}
\eht{\rho}\fht{D}_t \fht{c''d''} = & +\eht{c''\rho D_t d} - \eht{\rho}\fht{c''u''_n}\partial_n\fht{d} \nonumber \\ 
&+  \eht{d''\rho D_t c} - \eht{\rho}\fht{d''u''_n}\partial_n\fht{c} - \eht{\partial_n{\rho c''d'' u''_n}}
\label{eq:second-order-moments}
\end{align}
\noindent by substituting  $X_i$  for $c$ and  $u_r$ for $d$ and using the radial momentum equation
\begin{equation}
\rho D_{t} \big(u_{r}\big) =  \nabla \cdot {\bf \tau_r} - G_r^M - \partial_{r} P + \rho g_r
\end{equation}

\noindent where ${\bf \tau_r} = [\tau_{rr},\tau_{r \theta},\tau_{r \phi}]$ contains the radial components of the viscous stress tensor (not explicitly included in our simulation model), $G_r^M = -(\rho u_{\theta}^{2} - \tau_{\theta\theta})/r - (\rho u_{\phi}^{2} - \tau_{\phi\phi}) / r$ is a geometric term, $P$ is the pressure, and $g_r$ is the gravitational acceleration in the radial direction.

After averaging, we arrive at the flux evolution equation,
\begin{align}
\erho \fav{D}_t (f_i / \eht{\rho}) = &   -\nabla_r f_i^r  - f_i \partial_r \fht{u}_r - \fht{R}_{rr} \partial_r \fht{X}_i - \eht{X''_i} \partial_r \eht{P} \nonumber \\
&  - \eht{X''_i \partial_r P'} + \overline{u''_r \rho \dot{X}_i^{\rm nuc}} + {\mathcal G_i}  +  {\mathcal N_{fi}} \label{eq:rans_falpha}   
\end{align}

\noindent where $f_i^r = \eht{\rho}\fht{X''_i u''_r u''_r}$ is the radial component of the ``flux 
of the turbulent flux'' of element $i$. Here $-f_i \partial_r \fht{u}_r$ is a production term due to velocity effects controlled by the flux itself. The $-\fht{R}_{rr} \partial_r \fht{X}_i$ term is a production term which transports the flux from regions with higher $X_i$ to regions with lower $X_i$ or vice versa and is controlled by the Reynolds stress $\fht{R}_{rr} = \eht{\rho} \fht{u''_r u''_r}$. The terms  $-\eht{X''_i} \partial_r \eht{P}$ and $\eht{X''_i \partial_r P'}$ drive evolution of the flux in the presence of a pressure gradient and pressure fluctuations. The term $+\overline{u''_r \rho \dot{X}_i}$ drives the evolution of the flux by the net nuclear burning of element $i$. Finally, the term ${\mathcal G_i} = \eht{G_r^i} - \eht{X_i^{''} G_r^M}$, where $\eht{G_r^i}= -\eht{\rho X_i^{''} u_\theta^{''} u_\theta^{''}/r} - \eht{\rho X_i^{''} u_\phi^{''} u_\phi^{''}/r}$  mediates the production of $f_i$ due to centrifugal forces caused by horizontal (non-radial) motion of the flow. The equation contains additionally a term $\eht{X''_i \rho g_r}$ term, which we neglect due to gravitational acceleration $g_r$ being constant in  the simulation. 

The residuals of all these mean fields in the flux transport equation,
${\mathcal N_{fi}}$, represent numerical effects which we do
not calculate explicitly in RA-ILES, which has no explicit viscous (Navier-Stokes) term. 
Their magnitude is determined solely by the
implicit action of our numerical scheme at the sub-grid scale. The precise mathematical formulation of the term is following:
\begin{align}
{\mathcal N_{fi}} = &-\nabla_r (\eht{X''_i \tau_{rr}}) + \varepsilon_i^\tau + {\mathcal G_i^\tau} 
\end{align}
where
\begin{align}
\varepsilon_i^\tau = & - \eht{\tau_{rr}\partial_r X''_i} - \eht{\tau_{r\theta}(1/r)\partial_\theta X''_i} - \eht{\tau_{r\phi}(1/r\sin{\theta})\partial_\phi X''_i} \nonumber \\
{\mathcal G_i^\tau} = &+\overline{X_i''\tau_{\theta \theta} \tau_{\theta \theta}/r} + \overline{X_i''\tau_{\phi \phi} \tau_{\phi \phi}/r} \nonumber 
\end{align}
The term $-\nabla_r (\eht{X''_i \tau'_{rr}})$ is a
flux of a composition flux controlled by viscosity; $\varepsilon_i^\tau$ is a viscous dissipation of the turbulent flux and ${\mathcal G_i^\tau}$ is of geometric origin and contributes to flux production due to action of viscosity.


\subsection{Evolution Equation for Composition Variance} 
\label{sect:variance-evolution}

To derive a transport equation for the variance of composition fluctuations of an arbitrary element $i$, we use 
Eq.~\ref{eq:second-order-moments} and substitute $c$ for $d$:
\begin{align} 
\eht{\rho}\fht{D}_t \fht{c''c''} = & +2 \eht{c''\rho D_t c} - 2\eht{\rho}\fht{c''u''_n}\partial_n\fht{c} - \eht{\partial_n{\rho c''c'' u''_n}}.
\label{eq:variance}
\end{align}
\noindent Next we  
substitute a mass fraction of a given element $X_i$ for $c$ and find
\begin{align}
\eht{\rho} \fht{D}_t \sigma_i = & -\nabla_r f_i^\sigma - 2 f_i \partial_r \fht{X}_i + 2 \eht{X''_i \rho \dot{X}_i^{\rm nuc}} + {\mathcal N_{\sigma_i}}
\label{eq:sigma-transport}
\end{align}
\noindent where $\sigma_i = \fht{X''_iX''_i}$ is the variance of the mass fraction of species $i$; $f_i^\sigma = \eht{\rho X''_i X''_i u''_r}$ is the  turbulent flux of this variance; $2 f_i \partial_r \fht{X}_i$ is a down-gradient production/destruction; and $2 \eht{X''_i \rho \dot{X}_i^{\rm nuc}}$ is a variance source/sink due to nuclear burning. 

The term ${\mathcal N_{\sigma_i}}$ represents the dissipation of  composition variance due to the numerical scheme, which is discussed further in \S\ref{sect:results}.

\begin{figure*}
\includegraphics[width=1.0\hsize]{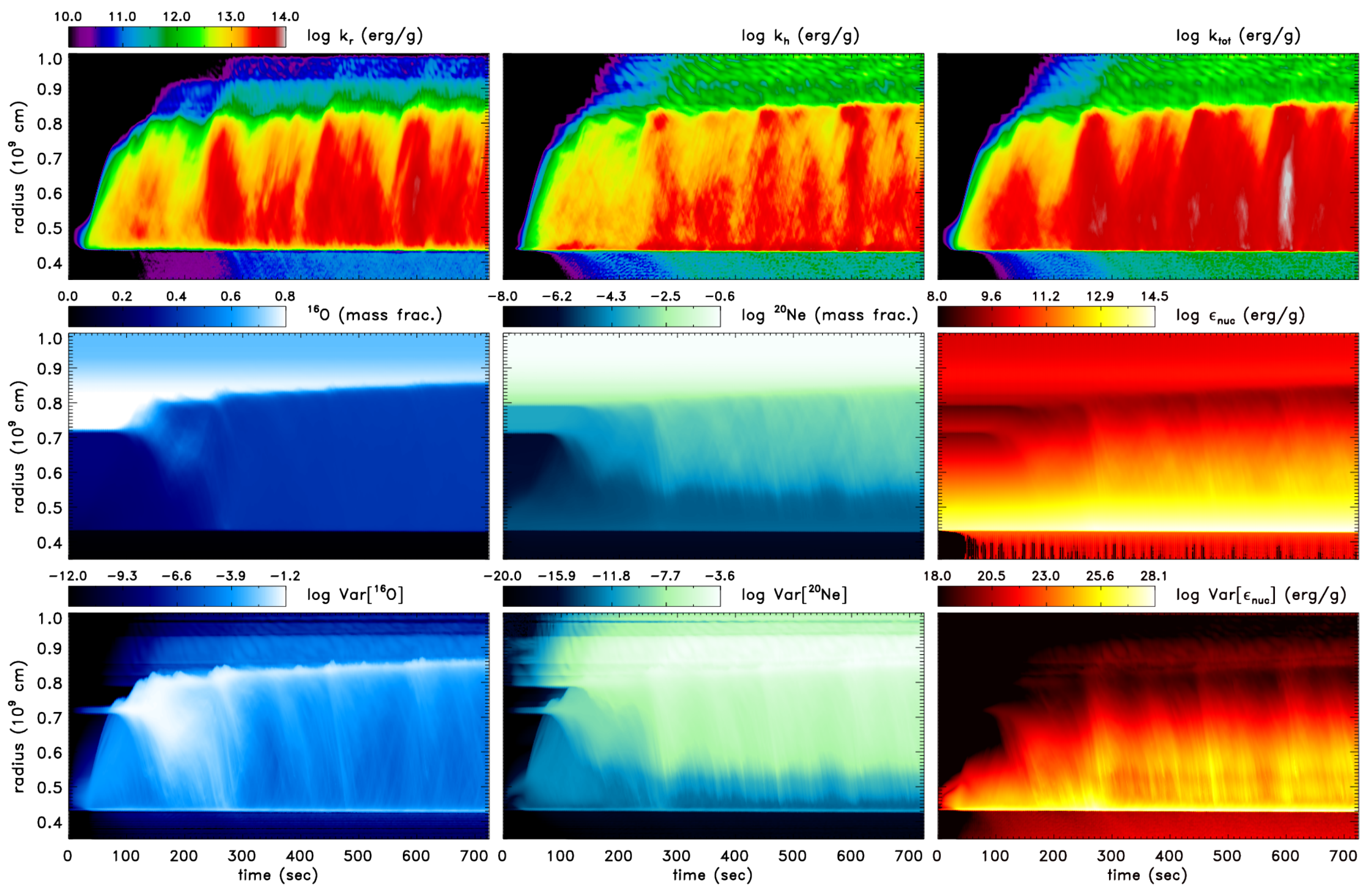}
\caption{\label{fig:kippenhahn} Space-time diagrams showing the evolution of various turbulence quantities for model {\sf ob.3d.B}.  From left to right, the {\em top row} shows: (left) radial turbulent kinetic energy (TKE); (middle) horizontal TKE; and (right) total TKE. From left to right, the {\em middle row} shows: (left) mean $^{16}$O abundance; (middle) base-10 logarithm of  mean $^{20}$Ne abundance; (right) mean net-nuclear energy generation rate. The {\em bottom row} shows, from left to right: (left) base-10 logarithm of $^{16}$O abundance variance; (middle) base-10 logarithm of $^{20}$Ne abundance variance; and (right) base-10 logarithm of net-nuclear energy generation variance.}
\end{figure*}

\section{Relevant timescales}
\label{sect:timescales}

The simulations are strongly dynamic, and may  be better understood  by comparison of several timescales, which we define here.
The convective turnover timescale is
\begin{align}
\tau_{\hbox{conv}} = 2 (r^c_\mathrm{t} - r^c_\mathrm{b}) / v_\mathrm{rms},
\end{align}
where $r^c_\mathrm{b}$ and $r^c_\mathrm{t}$ are the radii of the bottom and top convection boundaries, and $v_\mathrm{rms}$ is the rms of the velocity field in the convection zone.
The net nuclear (e-folding) burning timescale for element $i$ is
\begin{align}
\tau_{\rm \hbox{nuc}}^{i} = \fht{X}_i / \fav{\dot{X}}_i^{\rm nuc}.
\label{eq:net-nucl-timescale}
\end{align}
\par The nuclear (e-folding) burning timescale due to photo-disintegration of element $i$ is. 
\begin{align}
\tau_{\rm \hbox{nuc-phot}}^{i} = 1 / \lambda_{i} ,
\end{align}
and the nuclear (e-folding) timescale for burning element $i$ due to two-body reaction of $j$ and $k$ is
\begin{align}
\tau_{\rm \hbox{nuc-two}}^{i} = 1 / \left( \eht{\rho} \lambda_{jk} \fht{Y}_j \fht{Y}_k / \fht{Y}_i \right),
\end{align}
where $Y_i = X_i/A_i$ are molar abundances of species i, and $A_i$ is mean number of nucleons per isotope $i$. 
$\lambda_i$ and $\lambda_{jk}$ are nuclear reaction rates expressed as $N_A \langle\sigma v\rangle$, where $N_A$ is Avogadro's number, $\sigma$ is the reaction cross section, and $v$ the relative velocity of reactants \citep{Clayton1983,wda96}.

The (e-folding) transport timescale of element $i$ is
\begin{align}
\tau_{\hbox{tran}}^{i} = \fht{X}_i / (\nabla_r f_i / \overline{\rho});
\label{eq:transport-timescale}
\end{align}

\noindent and finally, the (e-folding) dissipation timescale of element $i$ is
\begin{align}
\tau_{\hbox{diss}}^{i} = \sigma_i / \varepsilon_i,
\label{eq:diss-timescale}
\end{align}
where $\varepsilon_i$ is the dissipation of variance $\sigma_i$ which we identify with the residual term $\mathcal N_{\sigma_i}$ defined in Eq.~\ref{eq:sigma-transport} above and discussed further in \S\ref{results:variance}.


\section{Results}
\label{sect:results}

To initiate convection in our 3D simulation we seed the initial
  hydrostatic model with small ($10^{-3}$) random perturbations in density
  and temperature \citep{MeakinArnett2007} in the unstable
  regions. Convection starts with an increase of turbulent kinetic energy
  near the base of the convection zone, similar to other qualitatively
  comparable cases, \eg \citet{Mocak2009,Stancliffe2011,Woodward2015}.  The
  velocities eventually grow from zero to $10^{-2}$ of sound speed, to
  become a self-consistent turbulent cascade, with final amplitudes that
  happen to be of the same order as MLT velocities because they involve the
  same buoyancy.

Since the 1D initial model does not contain sufficient dynamic information
for self-consistent quasi-static 3D convection, a readjustment occurs as a
self-consistent convective flow forms. This is usually referred to as the
``initial transient'' phase, and is a common feature of 3D stellar
hydrodynamics simulations; see \S\ref{initial-transients}. Even with
perfect hydrostatic matching of the initial model, a readjustment is
required -- 3D convection is different from 1D stellar convection theory
(e.g. \citealt{MeakinArnett2007,amvclm15}).

\subsection{Initial transients}\label{initial-transients}

For neutrino-cooled stages such as oxygen burning, convection is
  vigorous and turbulent \citep{MeakinArnett2007}. One-dimensional (1D)
  stellar evolutionary sequences use mixing-length theory (MLT), which
  assumes a specific average correlation between fluctuations in velocity
  and entropy; these correlations drive the fluxes crucial for thermal
  balance.  In contrast, 3D simulations must develop such correlations in a
  dynamically self-consistent way, to obtain the appropriate average. This
  takes a turnover time, after which the turbulent cascade can remove the
  excess entropy to obtain a balance {\em on average.}  Because mean
  velocities are dominated by {\em fluctuations} \citep{MeakinArnett2007},
  turbulent convection is made up of alternate bursts and pauses, occurring
  throughout the convection zone. It is intermittent; only the average is
  quasi-steady.  The 1D models cannot provide the dynamically consistent
  {\em phases} of the fluctuations which are crucial to accurately
  determine the net effect of cancellations, and for setting up a
  consistent 3D initial model.

In 3D stellar interior simulations, this fact, combined with the (generally
negligible) errors introduced by mapping a 1D model to a 3D space, leads to
an `initial transient' -- even if the most exacting hydrostatic balance is
enforced in the mean.

\subsection{Multi-layered convective-reactive burning}
\label{sect:convective-reactive-mixing}

During the initial transient phase (see \S\ref{initial-transients} above),
the convective plumes emerging from the oxygen burning shell quickly extend
into the neon burning region, which was previously separated by a very
narrow region of stability. The two shells merge within the first 100
seconds ($\sim$1 turnover) of the simulated star time. Soon a quasi
steady-state is established, when the total turbulent kinetic energy
density of the flow in the convection zone reaches roughly $10^{14}$ erg
g$^{-1}$, at around 300 seconds ($\sim$3 turnovers) of the simulation.

\begin{figure*}
\includegraphics[width=0.49\hsize]{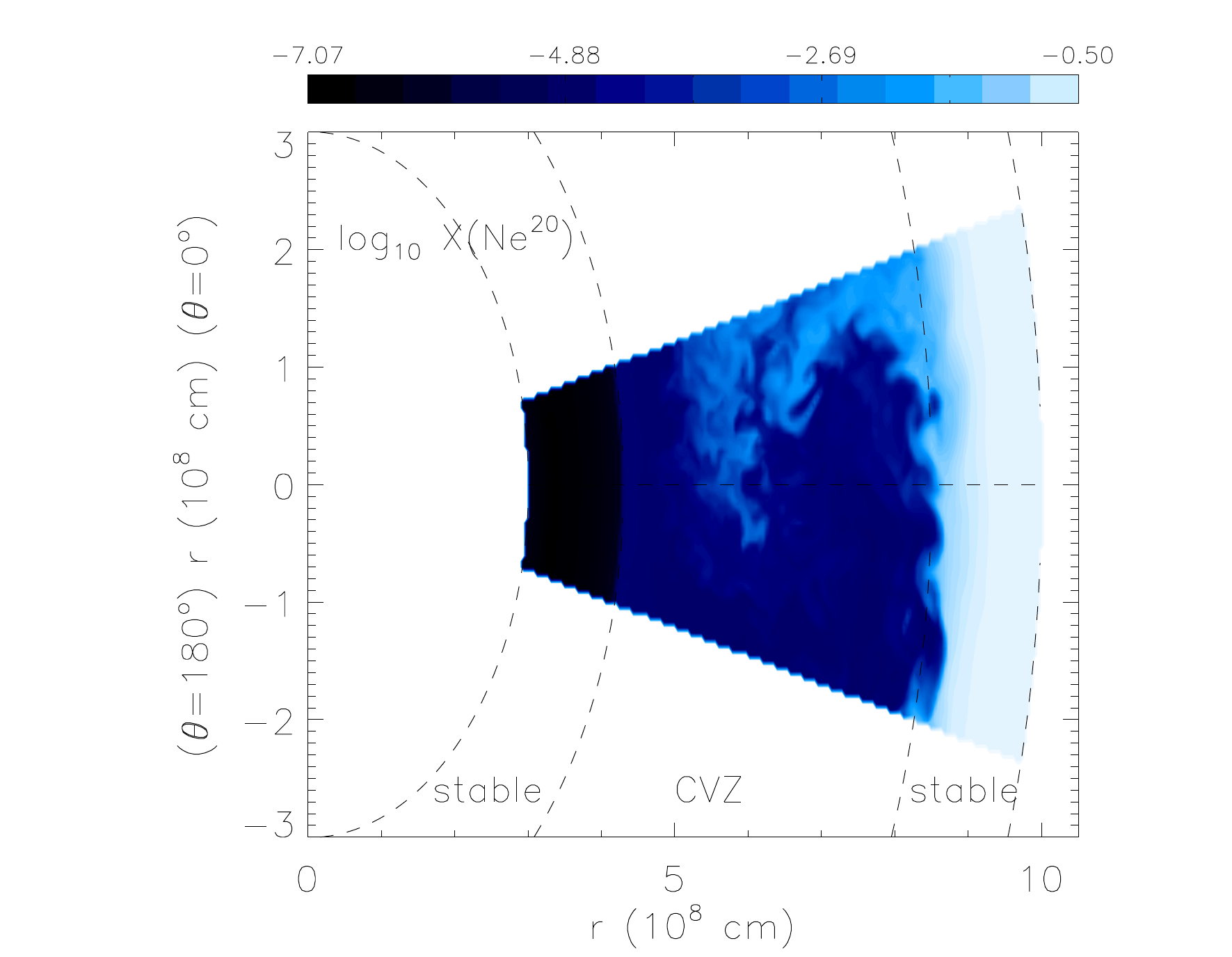}
\includegraphics[width=0.49\hsize]{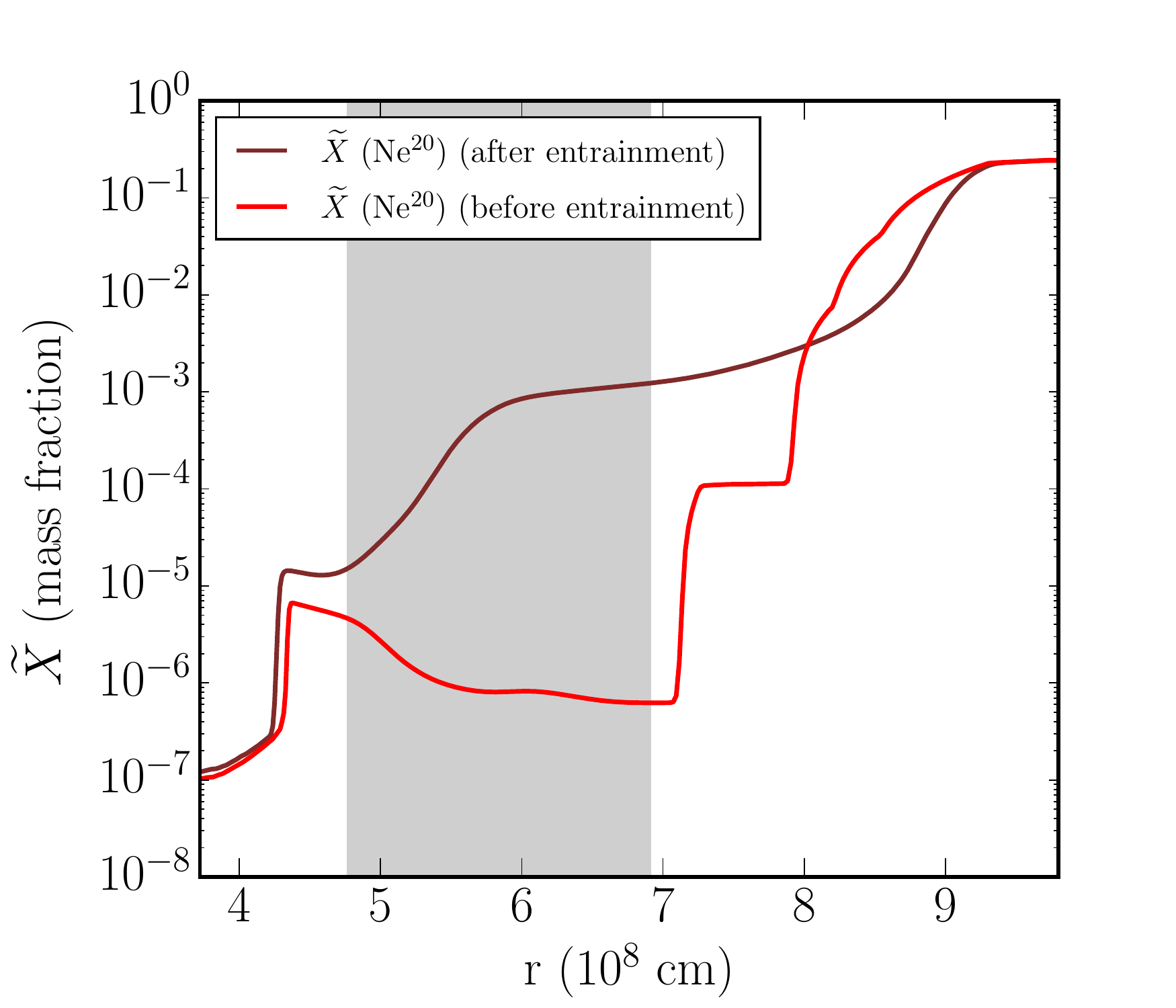}\\
\caption{{\texttt Left:} Snapshot of log$_{10}$ X($^{20}$Ne) in a meridional plane, where the inner dashed lines mark boundaries between convectively stable layers and convection zone (CVZ), and the outer dashed lines mark the edges of the computational domain. {\texttt Right:} Profiles of $^{20}$Ne mass fraction before (red) and after (brown) the entrainment event. The shaded vertical region highlights the area of significant nuclear changes in the $^{20}$Ne abundance, where its absolute rate of change due to nuclear burning $|\fht{\dot{X}}^{\rm nuc}_{ne20}|$ exceeds 2$\%$ of its maximum value. }
\label{fig:convective-reactive1} 
\end{figure*}

In a local 1D treatment of convection,  a linear criterion such as the Ledoux or Schwarzschild defines a region of stability which is capable of preventing mixing, regardless
of how weak the stable layer may be.  In reality, the
ability of a stable layer to survive against turbulent
convection depends on both the strength of the stabilizing gradient and the strength of the adjacent turbulence (the Richardson criterion; \cite{MeakinArnett2007,Cristini2017}). In this
particular model, the upper convective boundary was calculated by the Ledoux
criteria (at a radius around $7.2 \times 10^{8}$ cm), and was quickly overwhelmed by
the adjacent turbulent flow. The oxygen and neon-burning shells are
rapidly but gently\footnote{ No shocks or even strong pressure waves could be seen, only steady growth of turbulence.}  mixed together.

The onset of convection, the subsequent merging of the burning shells,
and the establishment of a quasi steady-state, is depicted in
Figure~\ref{fig:kippenhahn} as a group of ``Kippenhahn plots'', which are
space-time diagrams. Shown  
are the time evolution of (1)
the turbulent kinetic energy, (2) mean and variance of $^{16}$O and
$^{20}$Ne mass fractions, and (3) nuclear energy generation rates. The top row  
shows the time evolution of the turbulent kinetic energy
(TKE), including, from left to right, the radial, horizontal, and total
values. The kinetic energy at a given radius is calculated as half of the
variance of the velocity component in the horizontal plane, e.g., the
radial TKE $k_r$ is defined by $2 k_r(r) =\sigma_{u_r}^2 = \overline{(u_r -
  \overline{u}_r)^2 }$. The horizontal TKE is defined by $2 k_h =
\sigma_{u_{\theta}}^2 + \sigma_{u_{\phi}}^2$, and the total TKE by $k_{\rm
  tot} = k_r + k_h$.

The most prominent features of the quasi-steady state include a single
convection zone with two internal burning layers, as well as the
interplay of turbulent mixing and material entrainment. We describe these
features later for $^{16}$O and $^{20}$Ne 
nuclear burning\footnote{A more comprehensive presentation of the chemical
  element transport for all the remaining nuclear species in the model is
  presented in \citet{mmva14}.}. The net nuclear burning of other prominent
species in the convection zone (i.e., $^{12}$C or $^{28}$Si) are several orders
of magnitude smaller and are not discussed further here.

{\em This novel development is not presently captured in 1D stellar models: a robust current of entrained fuel is delivered from the neon-rich region to the hotter, deeper, oxygen rich layers,
without splitting the convective region. }
The process 
is most clearly seen for $^{20}$Ne which is being drawn into the oxygen burning
shell from the upper boundary. Figure~\ref{fig:convective-reactive1}  shows both ({\em left}) a cross sectional slice through the 3D simulation; as well as 
({\em right}) the radial profile of the Ne$^{20}$ mass fraction in both the initial model and
after a quasi-steady state has been achieved. The reason this phenomena has been missed in 1D models is that those models use MLT, which is a  local theory, and must be supplemented by ad hoc boundary conditions (Schwarzschild or Ledoux). The correct boundary conditions are nonlinear, dynamic and complex \ie not linear, static, and simple, as we shall see below.

\begin{figure*}
\includegraphics[width=0.49\hsize]{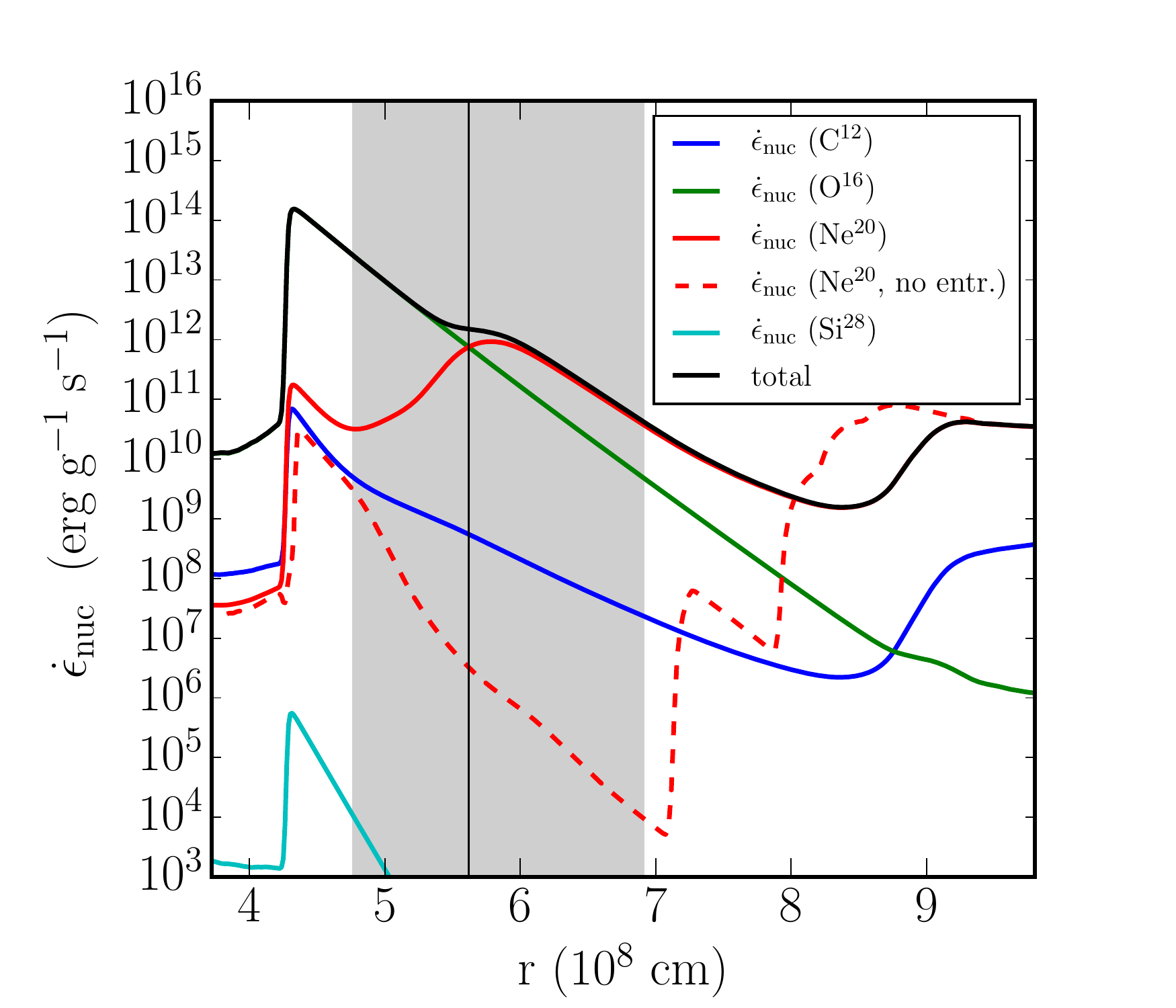}
\includegraphics[width=0.49\hsize]{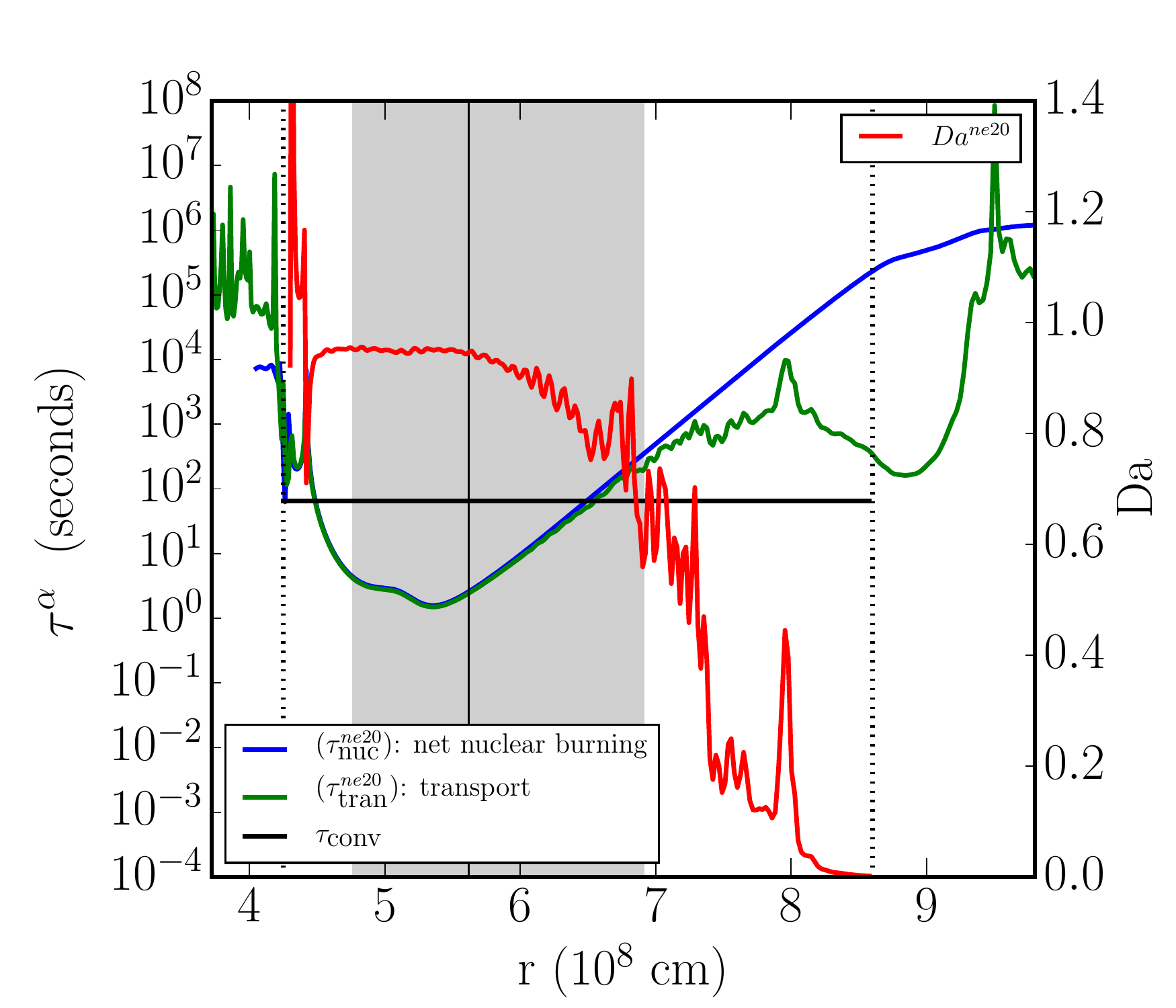}
\caption{{\texttt Left:} Nuclear energy production rates due to burning of $^{12}$C, $^{16}$O, $^{20}$Ne and $^{28}$Si, using the diagnostic approximations in Appendix \S\ref{nuclearapprox}. The vertical solid line marks the  maximum rate of change of neon density due to nuclear burning, at around $5.6 \times 10^{8}$ cm. The shaded vertical region highlights the area as described in Fig.\ref{fig:convective-reactive1}. The dashed line represents nuclear production rate due to $^{20}$Ne burning prior the entrainment event. {\texttt Right:} Damkh\"oler number $Da$ and the transport and net nuclear burning timescales $\tau$ for $^{20}$Ne as defined by Eq.\ref{eq:net-nucl-timescale}, Eq.\ref{eq:transport-timescale} and Eq.\ref{eq:damk}. The two vertical dotted lines mark convection zone boundaries. All curves are derived from mean fields calculated for central time at 1060 s of the simulation.} 
\label{fig:convective-reactive2} 
\end{figure*}

The nuclear energy
production increases in the vicinity of the temperature maximum by a factor
of 5. We find a secondary peak in the energy release at
$r\sim$\;5.7 $\times 10^{8}$ cm, where the main source of nuclear energy is
neon burning (Eq.~\ref{eq:neon-burning}). The contributions to the net nuclear energy generation
by different reactions are presented in the left panel of Figure~\ref{fig:convective-reactive2}. 

In the region of this secondary energy generation peak, the mass fraction of $^{20}$Ne is
maintained in a quasi-steady state by a balance between a net inflow from the overlying neon-rich 
stable layer and depletion by nuclear burning. This balance can be seen by comparing the timescales 
for the different processes involved, as shown in the right panel of Figure~\ref{fig:convective-reactive2}: 
the timescales for destruction of $^{20}$Ne by nuclear burning  (blue curve) and inflow by transport 
(green curve) match in the burning region. In contrast, the overlying layers are dominated by neon 
transport where its timescale is  $\sim$3 orders of magnitude shorter than that for depletion by 
nuclear burning.



A Damk\"ohler number, $Da$, may be defined as the ratio of transport to
nuclear burning timescales. Since we have more than one burning process,
multiple Damk\"ohler numbers should be defined, one for each reaction process $i$:

\begin{equation}
Da^i = \tau_{\hbox{tran}}^i / \tau^i_{\hbox{nuc}}
\label{eq:damk}
\end{equation}
For $Da \sim 1$, such that the transport timescale is comparable to
the burning timescale, we have the 
regime of convective-reactive
mixing \citep{Herwig2011} . In this regime it is 
important to
model the convection well, in order to predict the coupled mixing and
burning correctly. Such convective-reactive situations are likely to be common in stellar evolution --- if allowed by the evolutionary algorithm.  They might be common, for instance, in the early Universe in low-metallicity stars
\citep{Fujimoto2000,Schlattl2002}, or during dredge-up in AGB stars
\citep{Herwig2004,Goriely2004}.


The Damk\"ohler number varies across a convective region because of
the local variation in transport and nuclear timescales. 
The nuclear timescale depends on burning rates
which vary with fuel abundance, density, and
temperature, while the transport timescale
is determined by the topology of the velocity field. In the
upper layers of the convection zone in our simulation, the transport
timescale is much shorter than the nuclear burning timescales, so
$Da^{ne20} \ll 1$. There is much more neon entrained and transported
than nuclear burning is able to destroy. These timescales become close only
at around 7$\times 10^{8}$ cm, where effective nuclear destruction of
entrained $^{20}$Ne begins (see right-hand panel of
Fig.~\ref{fig:convective-reactive2}).



It is instructive to compare this 3D result to 1D 
simulations of
similar convective-reactive events.  One such class of 
calculations is the ingestion of protons into
He-buring convective regions (``proton ingestion episodes''). These
calculations result in a splitting of the He convection zone into two
parts (see e.g. \citealt{Herwig1999,Campbell2008}). The split occurs
because the large energy release from hydrogen burning creates a
temperature inversion at the point of peak H-burning luminosity, creating a
(formally) stable region just below. If time-dependent mixing is used in
the 1D calculation, this peak occurs at the point where the lifetime of a
proton against capture by $^{12}$C is just equal to the timescale of
convective mixing, i.e., $Da = 1$. In some studies it is explicitly enforced
that the hydrogen is mixed down to the point where $Da = 1$
(e.g. \citealt{Fujimoto2000}). 

In contrast, our 3D calculation shows that, in the neon ingestion case, the
$Da \sim 1$ condition defines a substantial burning region -- rather than a
single radial location. This suggests that a split of the convective zone
is less likely, as the burning is distributed rather than sharply
peaked. Such splitting\footnote{In the case of the star we have simulated,
  a splitting in 1D would be expected at a radius of $\sim 6.5 \times 10^8$
  cm. This is the location where $ \tau_{\hbox{conv}} = \tau_{\hbox{nuc}}$,
  i.e. the intersection of the blue and black lines in the right panel of
  Fig.~\ref{fig:convective-reactive2}. This is well above the location of
  the main burning in the 3D simulation (vertical line in
  Fig.~\ref{fig:convective-reactive2}).} is likely to be an artifact of the
1D formulation. In order for a radial velocity to go to zero in a
quasi-steady state, baryon conservation requires a finite horizontal
velocity \citep{amvclm15}, which in turn implies a shear instability and
probably mixing instead of splitting. This may be considered a boundary
issue, or ``overshooting''.

The (smooth) peak of nuclear energy release from neon burning is actually
located near the middle of the burning region, at around $5.7 \times 10^8$ cm
(Fig.~\ref{fig:convective-reactive2}). Neon mixes down still further, to $5
\times 10^8$ cm. As suggested by \citet{Herwig2011}, the width of
these regions must be due to a combination of factors, such as the
temperature sensitivity of the particular nuclear reactions, the range of
plume velocities, and the various mixing ratios in the plumes of (in our
case) neon-enriched material. In summary, the 3D simulations show that the
1D treatments of convective-reactive events miss some crucial elements that
could significantly change these phases.

{\em Our 3D results are qualitatively different from our 1D results for the same initial model, which we regard as a limitation in the 1D formulation.} At present 1D stellar evolution is apparently incomplete in its ability to represent spherically averaged 3D effects.
  The 1D stellar evolutionary sequence does not show a Ne-O
  shell merger, even at later times, and shows a failure just like the 1D results for ``splitting convection regions'' mentioned above. While there is some neon present in
  the O-burning convection zone of the 1D model
  (Fig.~\ref{fig:initial-model}), its nuclear energy production rate is
  negligible, being $\sim$6 orders of magnitude lower than that of oxygen (see
  bottom panel of Fig.~\ref{fig:convective-reactive2}; dashed-red curve). In contrast, the 3D
  model shows that the energy release from Ne burning dominates that of O
  in the top half of the convection zone. 
  
 Regardless of whatever the final answer may be for the evolution of Ne and O burning shells, the inconsistency of 1D and 3D models is  a cause to question present 1D evolutionary scenarios.
  
  
\begin{figure*}
\includegraphics[width=0.49\hsize]{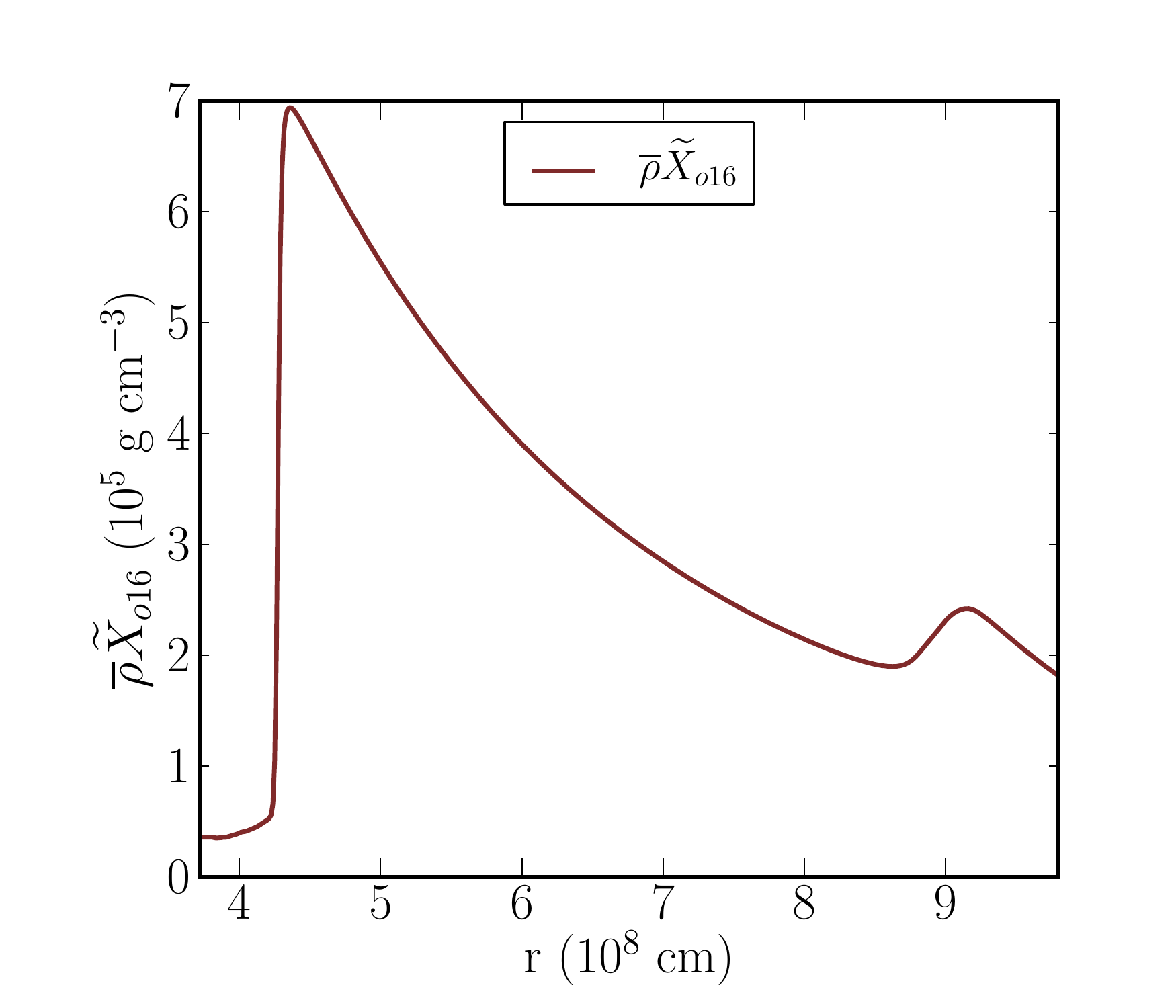}
\includegraphics[width=0.49\hsize]{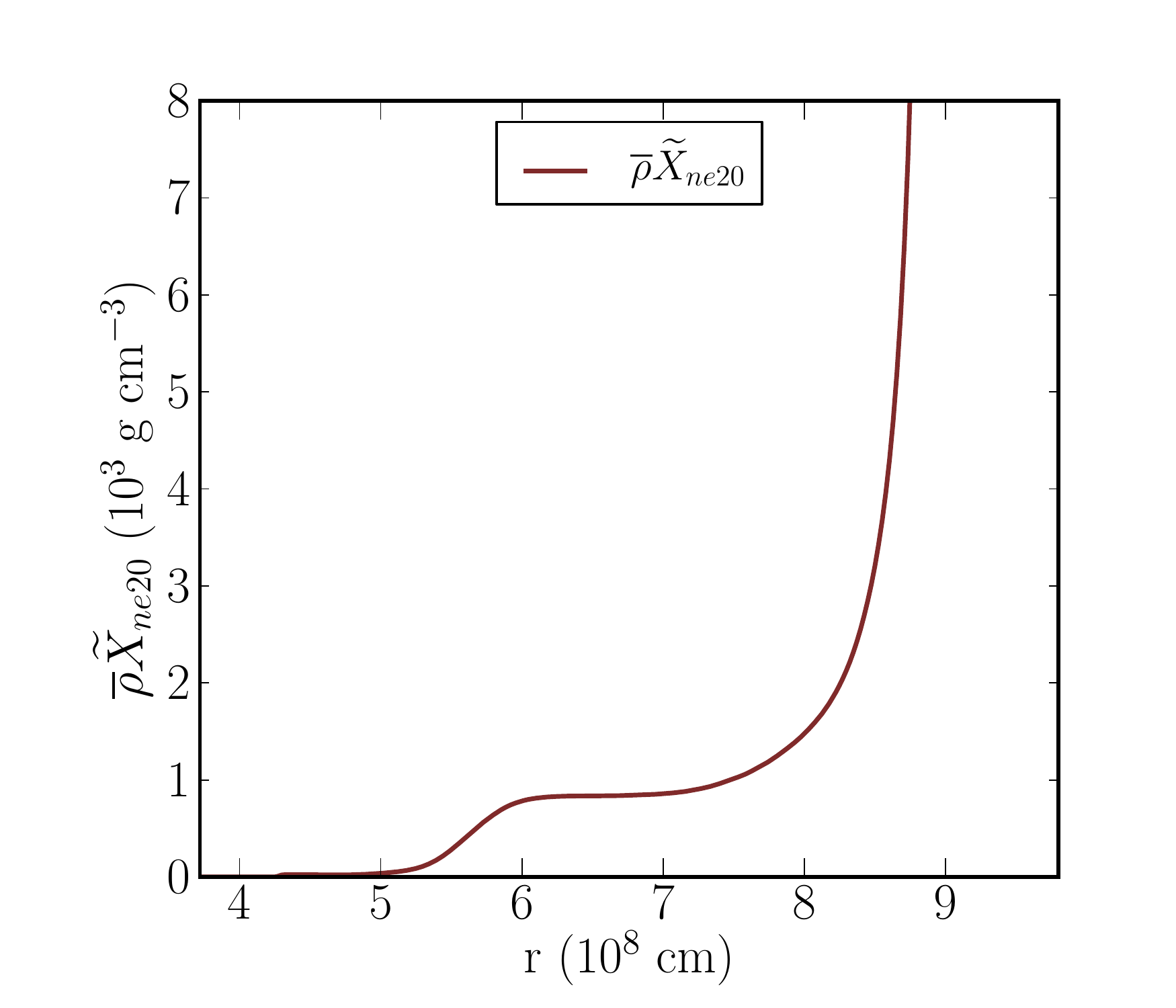}
\includegraphics[width=0.49\hsize]{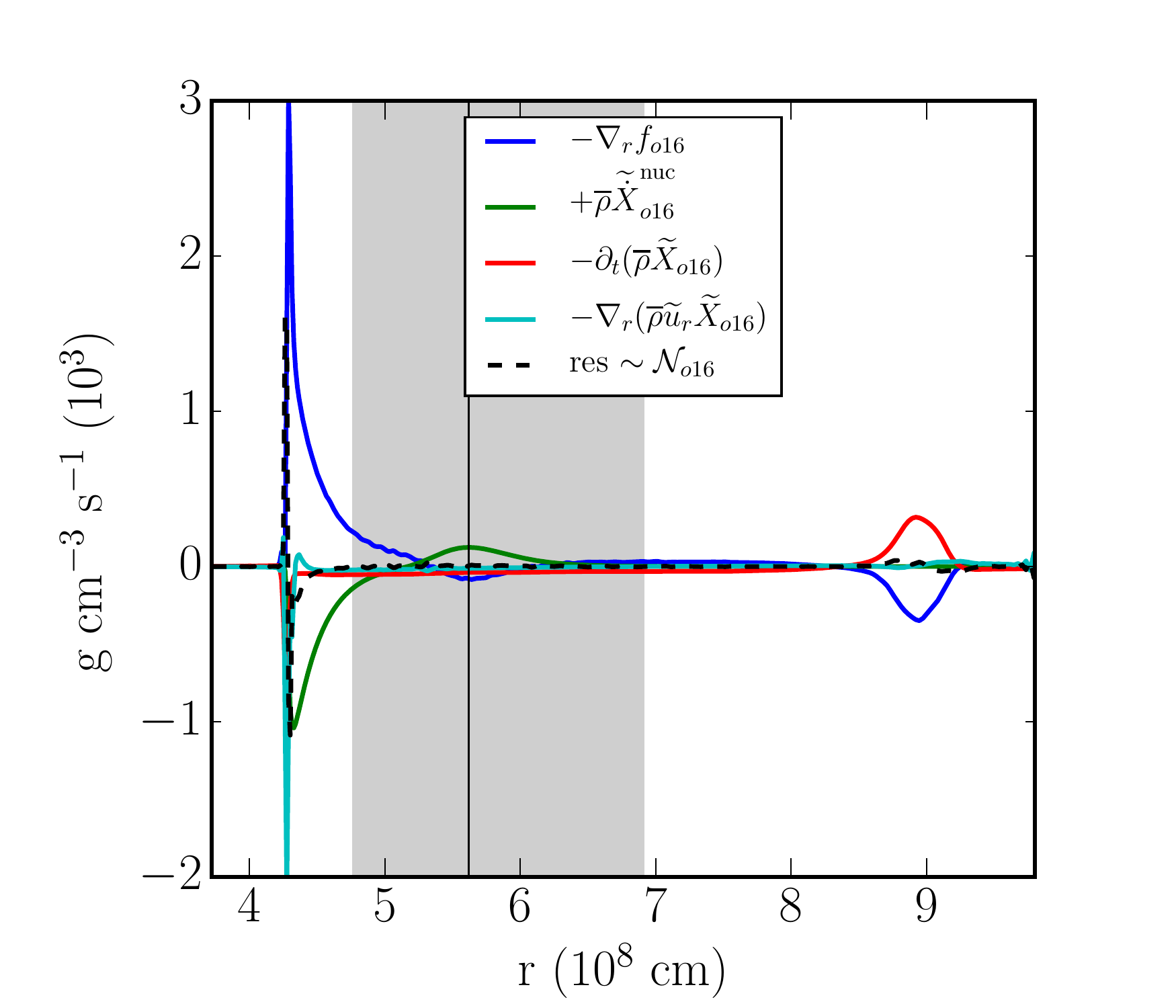}
\includegraphics[width=0.49\hsize]{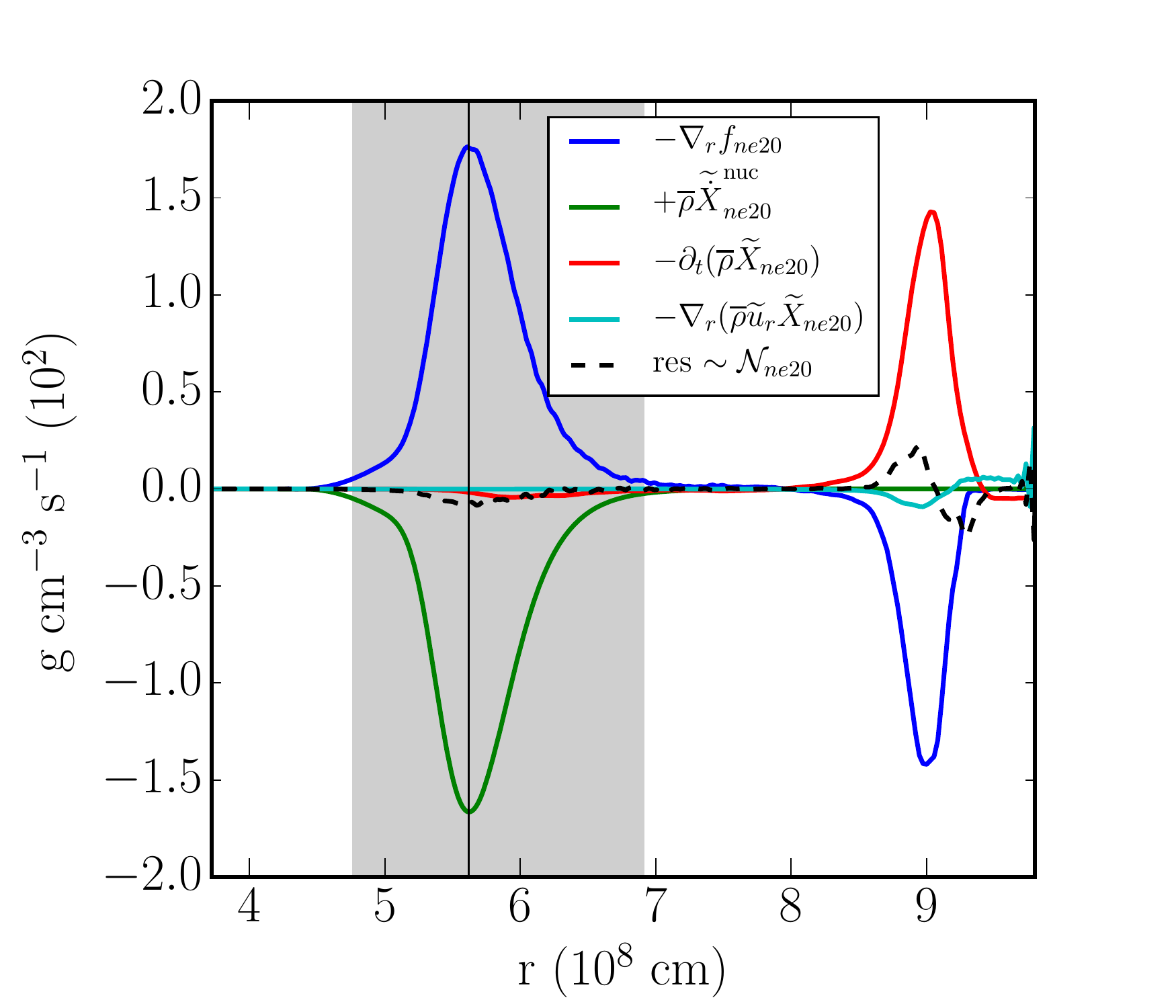}
\caption{{\texttt Top:} Profiles of $^{16}$O and $^{20}$Ne density mean fields. {\texttt Bottom:} $^{16}$O and $^{20}$Ne density transport equations. The shaded vertical region and vertical line as defined in Figure \ref{fig:convective-reactive2}.}
\label{fig:composition-transport}
\end{figure*}

\subsection{Mean-Field Analysis of 3D Simulation Data}

We use the three composition related mean-field RA-ILES equations to interpret our 3D
simulation data, which we depict graphically in 
Figs.~\ref{fig:composition-transport}, \ref{fig:flux-transport}, and \ref{fig:sigma-transport}. 
Each of these balance equations is discussed in the following 
subsection in turn, with a focus on a physical interpretation of each term in the equations.

\label{sect:mean-field-analysis}

\subsubsection{Mean Composition}\label{sec:meancomp}

The mean profiles of  $^{16}$O and $^{20}$Ne mass fraction are presented in the top row
of Fig.~\ref{fig:composition-transport}; while the bottom row of the same figure shows the individual
terms in the mean-field balance equation given by Eq.~\ref{eq:transport-comp} 
which underly these profiles. This figure shows that the composition profiles $\eht{\rho} \fht{X}_i$ are established
as an interplay between the material entrainment, turbulent mixing, and nuclear burning.
      
{\bf Time Dependence:} $\erho\fav{D}_t\fav{X}_i = \partial_t(\erho\fav{X}_i) + \nabla_r(\erho\fav{X}_i\fav{u}_r)$. While the convective layer is in a quasi-steady state, ongoing mixing at the upper boundary increases the depth of the convection zone over time. The signature of this mass entrainment appears in the Eulerian component of the time derivative, which is plotted as the red curve in Figure~\ref{fig:composition-transport}. By contrast, the change in the composition profile due to an overall expansion or contraction is negligible, as revealed by the turquoise curve. 

{\bf Turbulent Transport}: $-\nabla_r f_i = -\nabla_r(\erho \fav{u''_r X''_i})$.  The stable layer material which is mixed into the convective layer is transported by turbulent velocity fluctuations. This spatial redistribution of material is shown by the blue curve in Figure~\ref{fig:composition-transport} where we see that $^{16}$O and $^{20}$Ne are transported out of the boundary layer ({$r\sim 9 \times 10^8$ cm}) and into the lower nuclear burning regions of the convection zone ({$r < 7 \times 10^8$ cm}).  A simple balance is found: the transport out of the stable layer is closely matched by the time rate of change there; and at the other end, the rate at which material is being brought into the nuclear burning zone is  balanced by the rate it is being destroyed by nuclear reactions. 
 
  In the case of
  $^{20}$Ne, the  turbulent transport at the outer convective
  boundary shows depletion by downdrafts -- there is more $^{20}$Ne going
  out from the region than is going in. In contrast, the 
  transport is positive in the neon burning region below. This indicates
  that there is more neon going into this region than out, and the
  net input is equal to the $^{20}$Ne burned by nuclear reactions. The
  situation is almost identical for $^{16}$O, except there is additional
  transport of freshly produced oxygen in the neon burning layer due to its
  photodisentegration at $r\sim 5.7 \times 10^8$ cm.

{\bf Nuclear Burning}: $\erho\fav{\dot{X}}_i^{\rm nuc}$. The rate at which
the mean abundance of $^{20}$Ne is changed due to nuclear burning is
shown by the green curve in Figure~\ref{fig:composition-transport}. It
reveals a region of depletion near the lower part of the convection zone
where the temperature and densities rise above $\sim 1.9\times 10^9$ K and
$\sim 7\times10^5$ g cm$^{-3}$, respectively.  {\em The $^{20}$Ne burning rate is
  governed by the rate at which entrained material is transported into the burning region.  }

The width of the neon burning region is set at the upper end by the
reaction rates which are a function of the density and temperature, so that
burning does not begin until the material is transported below a radius of
$7 \times 10^8$ cm. At lower radii, the burning region is truncated by the
complete depletion of the $^{20}$Ne as it is being transported
downwards. The nearly symmetrical, gaussian shape is therefore a coincidence
of the near balance in timescales between these two competing processes: a
more vigorous turbulent transport, and hence a shorter convective mixing
timescale through the burning region, would result in a deeper lower
boundary consisting of fuel which was able to penetrate deeper before being
completely consumed.  The higher temperatures present near the base of the
oxygen-burning shell, however, ensures that the bulk of any entrained
$^{20}$Ne is consumed before traversing the entire convective layer,
resulting in a structure with ``stacked'' nuclear burning zones and a
stratified composition. The plateau in $\erho\fav{X}_{ne20}$, at  $r \sim 6.5\times 10^8$cm (see Fig.~\ref{fig:composition-transport}), 
 reflects this balance between entrainment and burning.

The transport of $^{16}$O in the upper layers of the convection zone
shows similar properties to that of $^{20}$Ne, with
oxygen being pulled down into the convection zone from the upper stable layer by turbulent velocity
fluctuations. However, in layers where Ne burning dominates,
$^{16}$O is produced through photo-disentegration:
$\rm ^{20}Ne(\gamma,^{16}O)^{24}Mg$ (see lower left panel in Fig.~\ref{fig:composition-transport}).
Toward smaller radii, 
oxygen depletion continues to increase due to nuclear fusion.

{ \bf Numerical Residual:} $\mathcal{N}_i$. Unlike traditional solvers of the Navier-Stokes equations, a conservative hydrodynamics
algorithm such as PPM results in errors in conservation laws with magnitudes comparable to 
machine precision. Therefore, the residuals found in the balance equations are due to the much larger effects -- and can be understood as loss of information at the grid level (dissipation and diffusion).

 A quantitative measure of this numerical residual is shown by the dashed black curve in
Figure~\ref{fig:composition-transport}. The impact that it has on the mean flow can be quantified by comparing the amplitude of
this curve with the other dominant physical processes controlling the time
evolution. In comparison to nuclear burning and turbulent transport, which
are the dominant processes, the numerical flux is relatively
  small over the majority of the domain. However, it does make an important contribution 
  in narrow regions around both the upper and lower convective boundaries
  (Fig.~\ref{fig:composition-transport}). Here turbulent boundary layers develop (\cite{llfm1959}, \S44).
  As expected, this flux is present
  exactly where the mean composition fields possess steep gradients. The
  steady presence of internal wave fields in and around the convective
  boundaries ensures that the material fields are constantly in motion
  against the computational mesh, thereby weakening the gradients. The
  strength of this numerical flux is seen to diminish with finer
  resolution as gradients are better resolved \citep{amvclm15,Cristini2017}.

\subsubsection{Turbulent Composition Flux}\label{sec:compflux}

\begin{figure*}
\includegraphics[width=0.49\hsize]{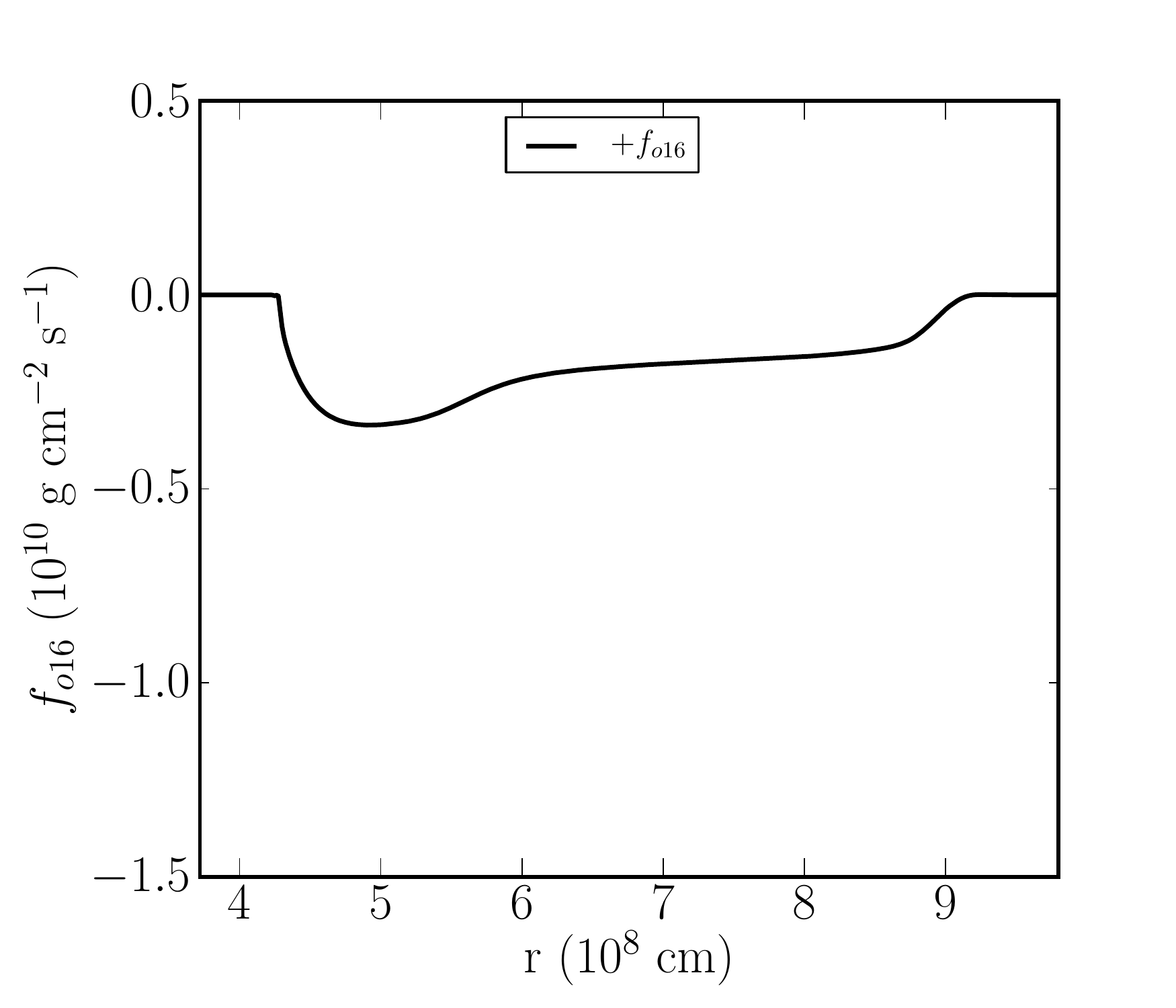}
\includegraphics[width=0.49\hsize]{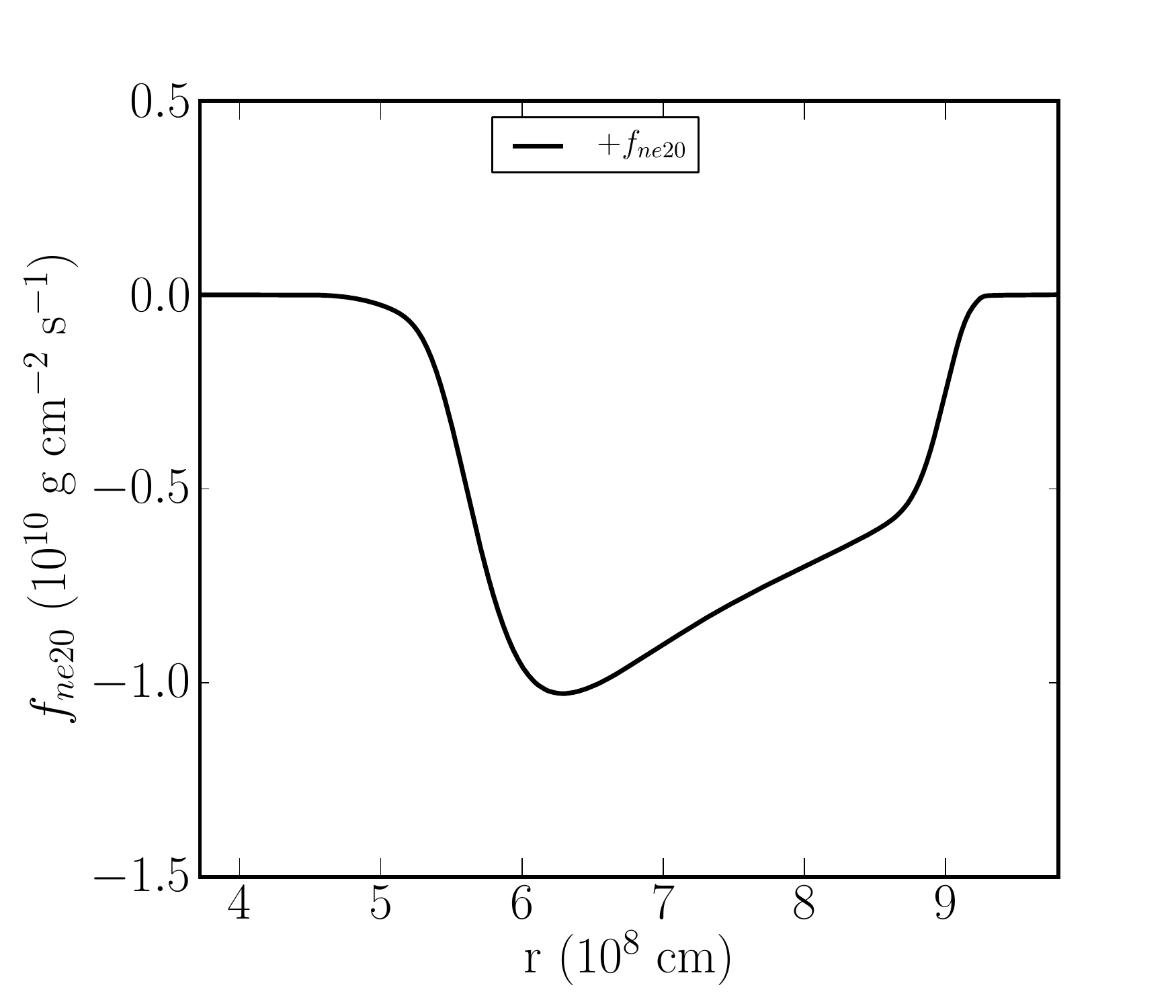}

\includegraphics[width=0.49\hsize]{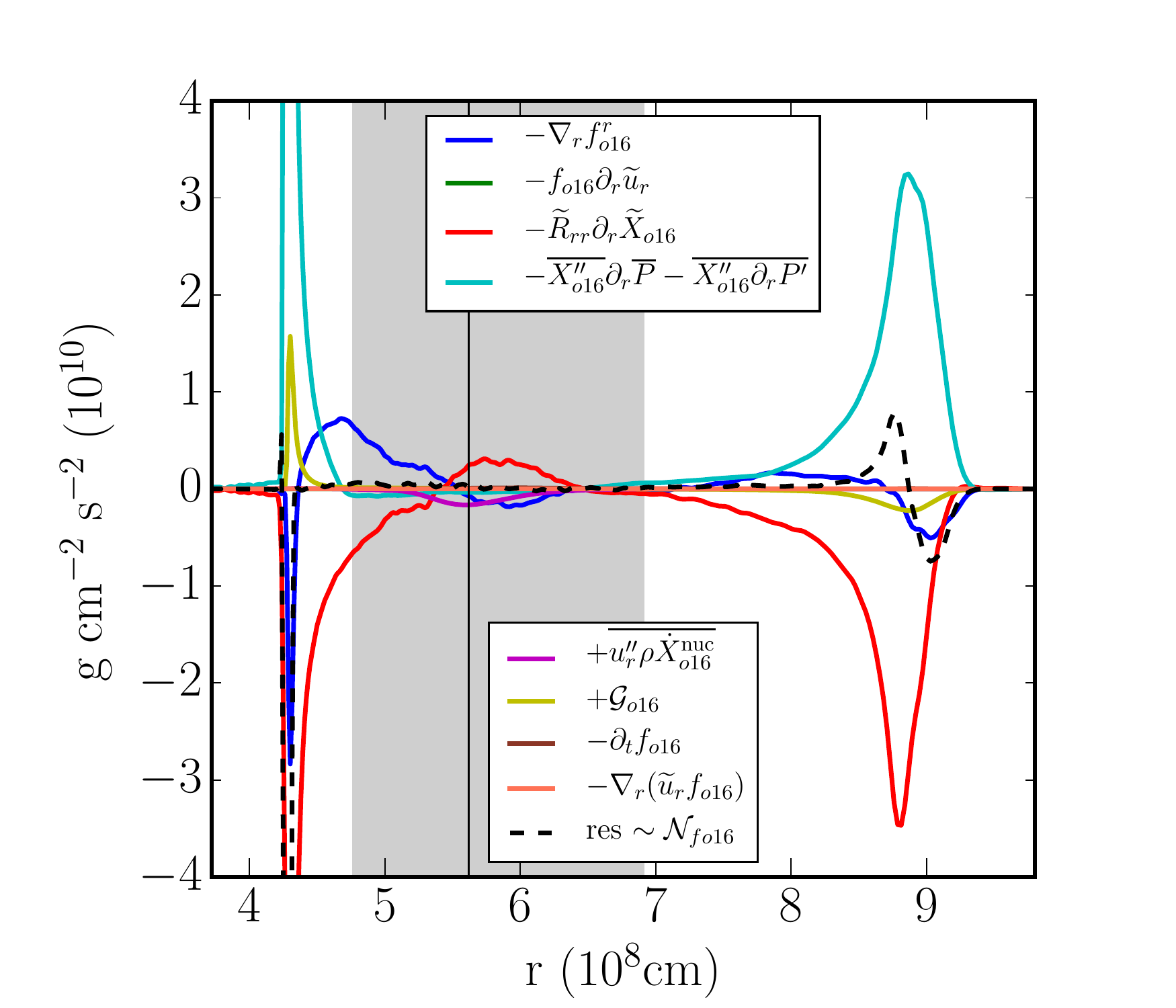}
\includegraphics[width=0.49\hsize]{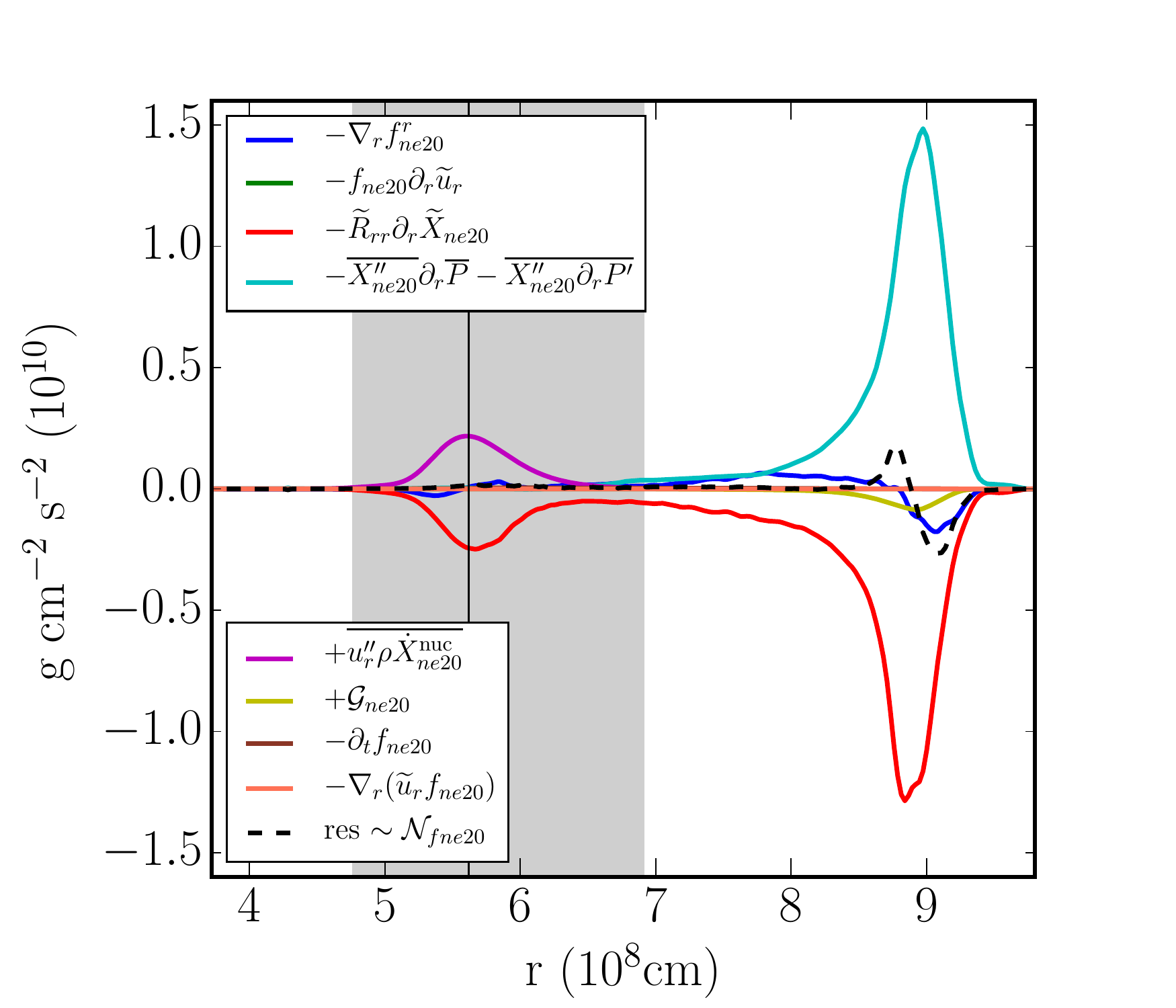}
\caption{{\texttt Top:} Profiles $^{16}$O and $^{20}$Ne turbulent flux mean fields. {\texttt Bottom:} $^{16}$O and $^{20}$Ne turbulent flux mean fields transport equations. The shaded vertical region and vertical line as defined in Figure \ref{fig:convective-reactive2}.}
\label{fig:flux-transport}
\end{figure*}

The turbulent composition flux $f_i$ represents the rate at which matter of species $i$ is transported across a given spherical shell. Flux profiles and balance equation terms for the flux evolution equations for $^{16}$O and $^{20}$Ne are presented in Figure \ref{fig:flux-transport}. The negative values of the fluxes indicate that they are oriented towards the stellar center, which is consistent with the picture of entrainment from upper stable layers dominating the mixing of new material into the convection zone. 
 
{\bf Time Dependence:} $\erho\fav{D}_t (f_i/\erho) =
  \partial_t(f_{i})$ + $\nabla_r(f_{i}\fav{u}_r)$. \\ The rate of change of the flux, at a given spatial location 
  and within a fixed mass shell (i.e., at a fixed Lagrangian coordinate) -- shown by the orange and brown curves 
  in Fig.~\ref{fig:flux-transport} -- are both found to be small compared to the other terms determining the flux profile, 
  indicating a quasi-steady balance. The expansion velocities in the model are only
  $\sim10^2 \cms$, which is 5 orders of magnitude smaller than the r.m.s. of the
  turbulent velocity field ($\sim 10^{7} \cms$). The quasi-steady flux of $^{16}$O and
  $^{20}$Ne is here due to a balance
  between nuclear burning, turbulent transport, and pressure-force and centrifugal-force coupling.

{\bf Turbulent Production:} $-\fav{R}_{rr}\partial_r\fav{X}_i$. The presence of radial velocity fluctuations in regions where there is a mean radial composition gradient leads to the production of a turbulent mass flux. This process involves the radial component  of the turbulent kinetic energy $k_r = \fav{R}_{rr}/2$ and the mean composition gradient for isotope $i$, given by $\partial_r\fav{X}_i$. In the case of $^{20}$Ne, where the mean abundance gradient is everywhere positive, the presence of turbulent velocity fluctuations leads to a {\em downward directed} mass flux, and hence has a {\em negative sign}. This term is shown by the red curve in Figure~\ref{fig:flux-transport}. The situation is similar for $^{16}$O with the exception of (1) the neon burning region -- where the oxygen abundance locally increases due to photo-disintegration of neon -- giving rise to a contribution to the outward-directed composition flux; and (2) the region around the lower boundary, where oxygen burning produces a very steep gradient and hence a large contribution to the downward-directed flux component. 

{\bf Pressure-force Coupling:} $-\left(\eht{X''_{i}}\partial_r \eht{p} + \eht{X''_i\partial_r p'} \right)$. This process is localized near the upper and lower convective boundaries where it peaks and drops off on either side of the interface.  These terms are shown by the turquoise colored curve in Figure~\ref{fig:flux-transport}. On the convection zone side of each boundary, the overall strength of this effect drops as the buoyancy diminishes relative to the velocity fluctuations; on the stable side of each boundary, the strength of buoyancy is set by the entropy stratification while the turbulent velocity fluctuations decay with distance from the boundary.

Material which resides in the stable layer is buoyant relative to material in the adjacent convection zones by virtue of its relative mean molecular weight  and entropy: in the upper stable layer, the material is lighter per ion and is in a higher entropy configuration. Therefore, as material is drawn down into the convection zone by turbulent velocity fluctuations its downward motion is resisted by this buoyancy.  As the freshly entrained stable layer material is dredged deeper into the convection zone, it is continually being mixed with its surroundings by turbulence which reduces its relative buoyancy. By the time this newly ingested material reaches the far end of the convection zone the relative buoyancy affecting its flux is negligible. 

The part of this term that reads $-\eht{X''_{i}}\partial_r \eht{p}$ can be rewritten to highlight its relationship to buoyancy: using the identity $\eht{X''_i} = -\eht{\rho X'_i}/\erho$ and hydrostatic equilibrium $\partial_r\eht{p} = - \erho\fav{g}$ we find 
\begin{align}
-\eht{X''_{i}}\partial_r \eht{p}  = \eht{X'_i\rho'}\fav{g}.
\end{align}
The term $-\eht{X''_i\partial_r p'}$ is related to pressure fluctuations and is of a dynamic rather than a thermodynamic origin.  In the anelastic approximation, the pressure fluctuations are governed by a Poisson equation
\begin{align}
\Delta p'  = -\nabla:(\rho_0 \vec{u}'\oplus\vec{u}') - \nabla\cdot(\rho'\vec{g}).
\end{align}
In stably stratified layers the buoyancy component dominates; in the well-mixed turbulent layer the advective term dominates \citep{VialletMeakin2013}; and the partially mixed boundary layer is a region of transition between the two.

{\bf Centrifugal-force Coupling:} $\mathcal{G}_i$. This term is driven by horizontal flow produced by the deflection of plumes at the inner boundaries of the convection zones and provides a radially directed acceleration to the matter with which it is coupled. At the upper boundary, the plumes which ultimately become horizontal flows carry material which is primarily representative of the mean composition of the convection zones, and hence will be depleted in those isotopes which are being actively entrained at the upper boundary. Therefore, the net effect of this process will be a net {\em negative} contribution to the entrained mass flux. Since we are considering downward directed fluxes of material being entrained at the upper boundary, this term will therefore act as a {\em source} for this entrained matter flux.  This term is plotted as the yellow curve in Figure~\ref{fig:flux-transport}.

A similar, but mirrored effect, takes place at the lower boundary for $^{16}$O.

{\bf Nuclear burning:} $\eht{u''_r \rho\dot{X}_{i}}$. As the $^{20}$Ne approaches the bottom of the convection zone it begins to undergo nuclear burning which diminish fluctuations in its abundance, and therefore reduces the downward-directed mass flux.  This process is represented by the magenta curve in Figure~\ref{fig:flux-transport} where it can be seen to operate in the neon burning region. This same process also operates for the $^{16}$O flux, but with the opposite sign.

{\bf Turbulent Transport:} $-\nabla_r{f^r_i} = -\nabla_r\left(\erho\fav{u''_ru''_rX''_i} \right)$.  In the same manner in which velocity fluctuations act to redistribute Reynolds stress, or its trace, the turbulent kinetic energy $k=Tr(\erho\fav{u''_iu''_j})$; velocity fluctuations also redistribute compositional flux ($f_i = \erho\fav{u''_rX''_i}$) in space.  Although not a dominant effect in controlling the radial profile of $^{20}$Ne flux, it does transport some of the downwardly directed flux from the middle of the convection zone to the upper stable layer and the region just below the $^{20}$Ne burning zone.  This spatial redistribution can be seen in Figure~\ref{fig:flux-transport} where it is represented by the darker blue curve.

{\bf Numerical Residual:} $\mathcal{N}_{fi}$. The implicit action of
turbulent cascade on the turbulent composition fluxes is apparent at the convective
boundaries.

\subsubsection{Composition Variance and its Dissipation Rate}
\label{results:variance}

\begin{figure*}
\includegraphics[width=0.49\hsize]{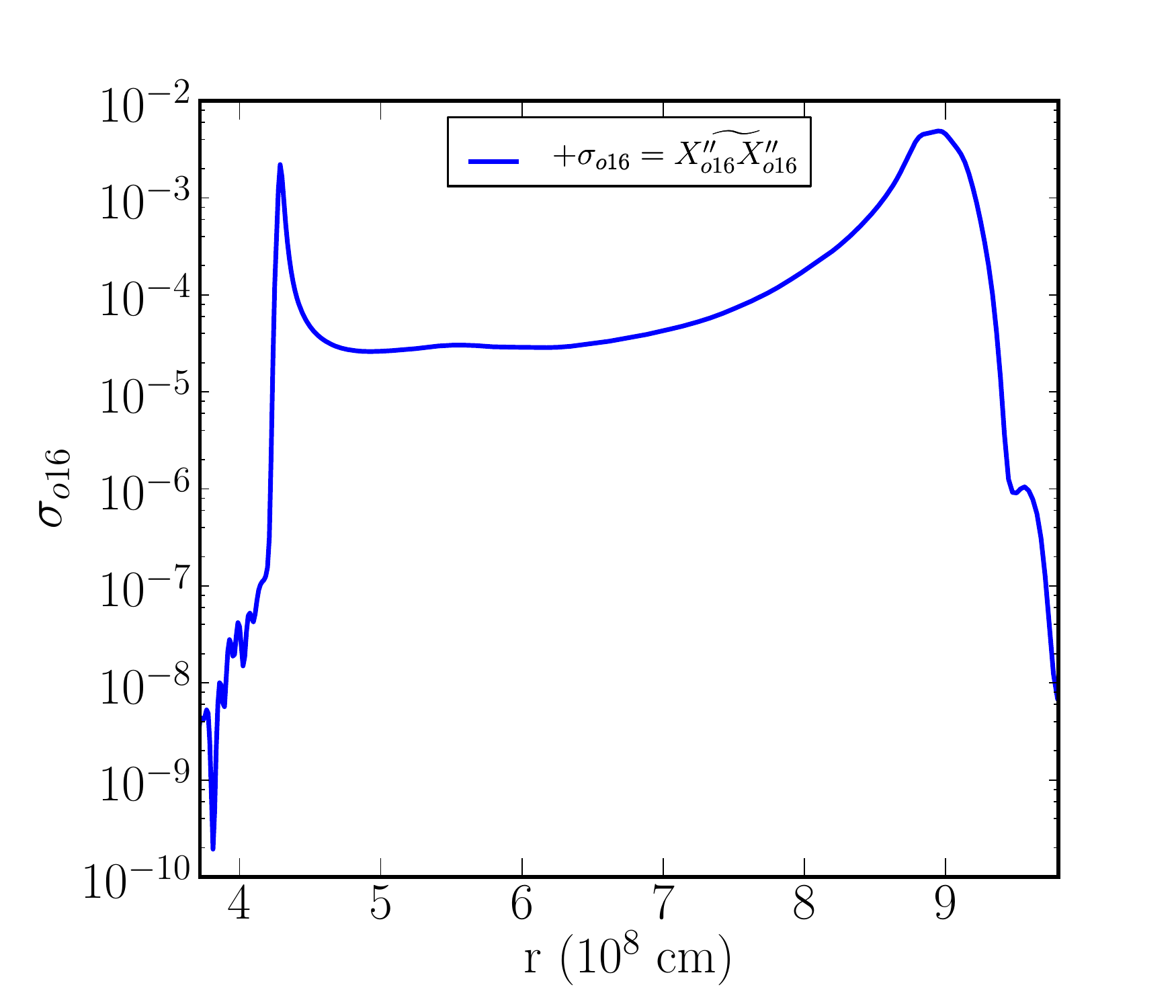}
\includegraphics[width=0.49\hsize]{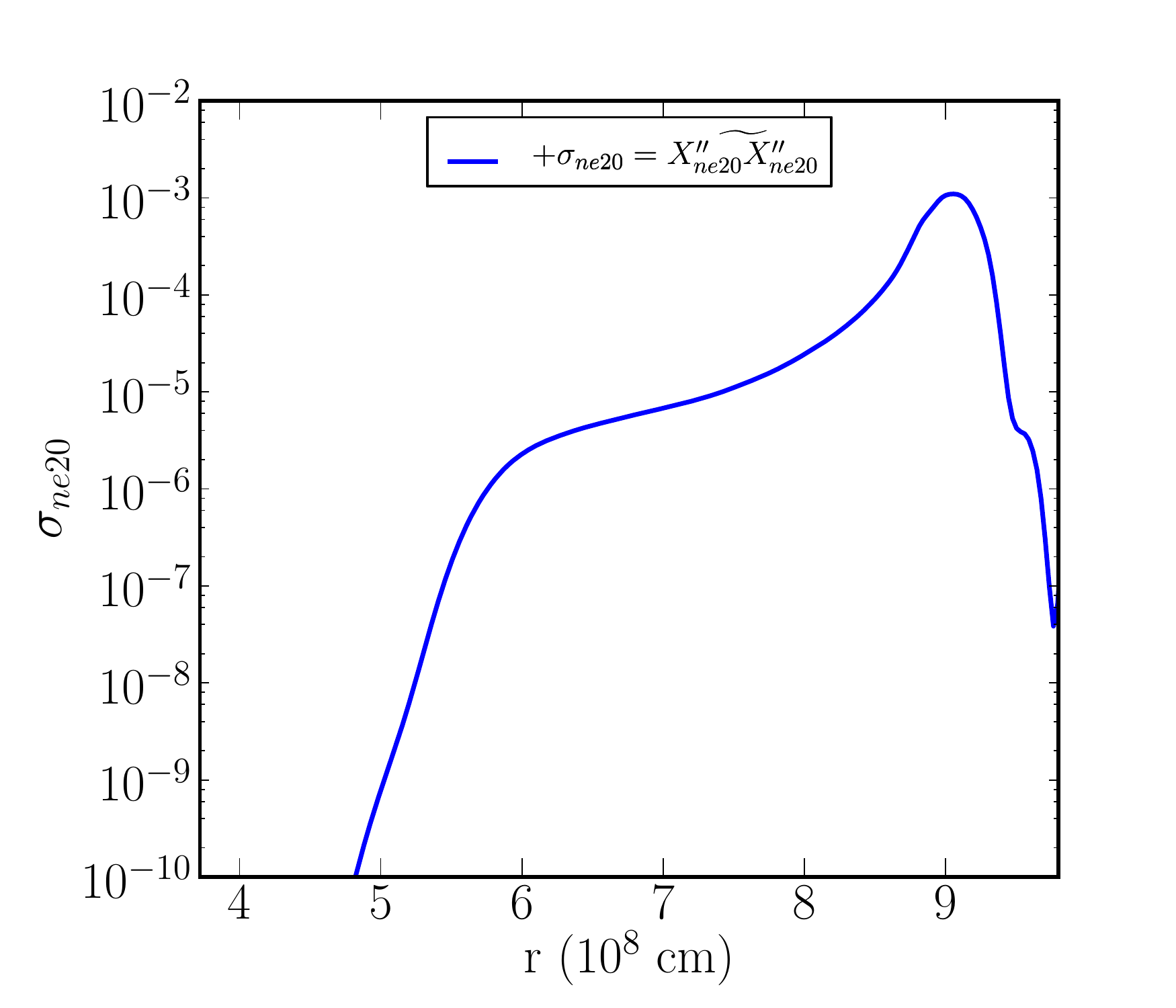} \\

\includegraphics[width=0.49\hsize]{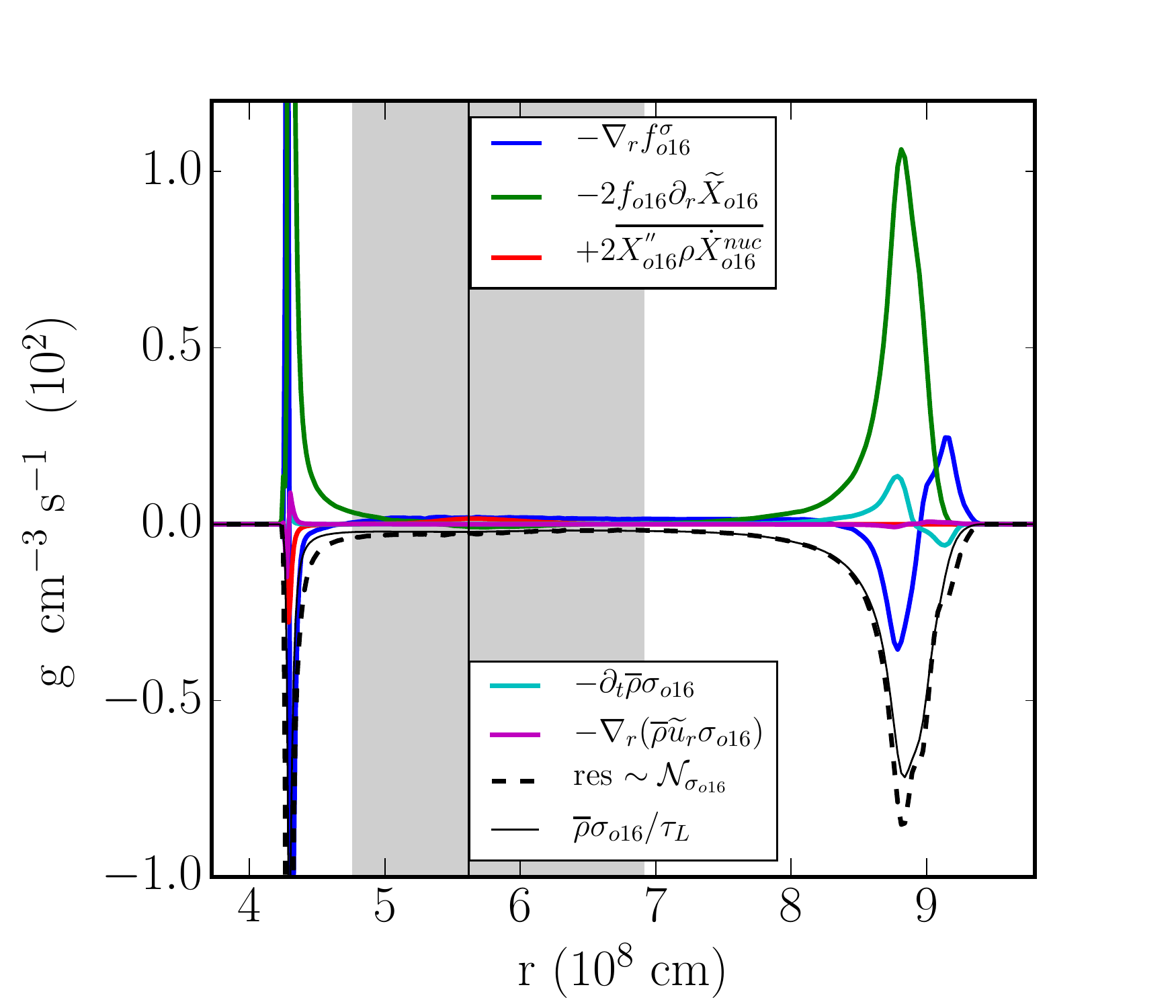}
\includegraphics[width=0.49\hsize]{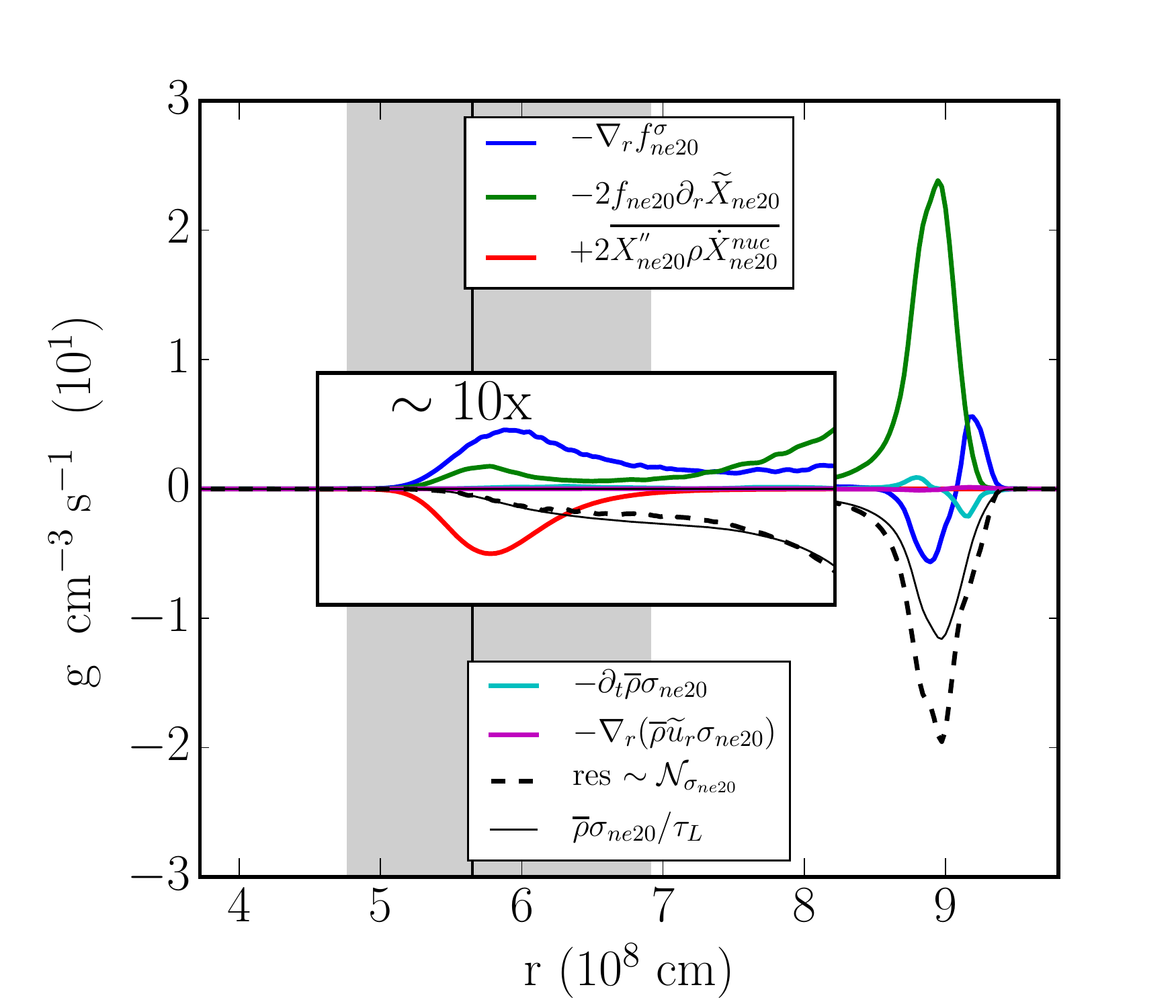} \\

\caption{{\texttt Top:} Profiles of $^{16}$O and $^{20}$Ne variances. {\texttt Bottom:} $^{16}$O and $^{20}$Ne variance transport equations. The shaded vertical region and vertical line as defined in Figure \ref{fig:convective-reactive2}.}
\label{fig:sigma-transport}
\end{figure*}

Composition variance $\sigma_i$ is a measure of the amplitudes of the
composition fluctuations at a given radial location in the stellar interior. 
It is typically largest at convective boundaries
(where the gradients are steepest), due to the oscillatory
distortion of the boundary by gravity waves as well as the presence of turbulent
entrainment of material into the convection zone. This  $\sigma_i$ and the terms in the balance equation for this
mean-field variable is shown in Figure \ref{fig:sigma-transport}. 
The $^{16}$O variance distribution shows maximum values
at both convective boundaries, whereas $^{20}$Ne peaks only at the upper
boundary  due to a lack of neon at deeper layers due to depletion by nuclear burning 
and turbulent mixing.

Dissipation is an important feature of turbulence.  In the ILES
approach, the Euler equations are solved so that 
dissipation is implicitly included; in our case,
the numerical algorithm provides an effective damping at the dissipation (sub-grid) scale.
The rate at which composition fluctuations are destroyed by this process can
be expressed by a dissipation timescale, as in Eq.~\ref{eq:diss-timescale}. We associate this timescale with the residual of the transport equation of variance
(Eq.~\ref{eq:sigma-transport}) since it characterizes the magnitude of
numerical dissipation. We discuss this important feature further in the paragraph below,
under the subheading {\em Numerical Residual}.

{\bf Time Dependence:} $\erho\fav{D}_t\sigma_i = \partial_t(\erho\sigma_i) + \nabla_r(\eht{\rho} \fav{u}_r \sigma_i)$. The variance is in a steady state throughout most of the computational domain but does show a sign of the convective boundary mixing at the upper boundary. The signature of this mass entrainment appears in the Eulerian time derivative which is shown by the turquoise curve in Figure~\ref{fig:sigma-transport} -- while the overall effect of background expansion on the mean variance field is negligible, as shown by the magenta curve.

The sine wave-like shape of the time dependence is due to the outward
migration of the variance peak following the mass entrainment (note the sign of the terms plotted as indicated in the figure legend). 

{\bf Turbulent Production:} $-2f_i\partial_r\fav{X}_i$. The primary source of variance in the composition field is turbulent production, which is analogous to the source of turbulent flux discussed above. In this case, variance is produced when a turbulent mass flux is present in regions where there is a mean composition gradient. Physically, the growth of variance can be thought of as the coupling of the composition fluctuation associated with the turbulent flux to the newly created composition fluctuation which is generated by the associated velocity fluctuation. 

In the case of the $^{20}$Ne, the net downward directed composition flux works against a positive gradient in the $^{20}$Ne abundance (see Fig.~\ref{fig:convective-reactive1}), thus amplifying the local fluctuations and resulting in an overall positive increase in the mean variance.  Production is seen to dominate in and around the upper boundary, then drops off with depth into the convection zone, until a small peak in the $^{20}$Ne burning region where again the mean composition gradients are very steep -- in this case due to the nuclear depletion of $^{20}$Ne. This term is shown by the green curve in Figure~\ref{fig:sigma-transport}.  The situation for $^{16}$O is similar, but has significant contributions at both boundaries.

{\bf Turbulent Transport:} $-\nabla_r f_{i}^\sigma = -\nabla_r\left(\erho\fav{u''_rX''_iX''_i} \right)$. Variance is transported by the variance flux $f_{i}^\sigma$ which is the correlation of a velocity fluctuation with the square of the composition fluctuation. As with all turbulent transport, it is a conservative process and neither creates nor destroys variance but simply redistributes it spatially. The redistribution of $^{20}$Ne has a relatively simple behavior in our 3D models: the variance that is generated within the upper boundary by turbulent production is moved both outward, to a region lying just above the main region of turbulent production; as well as being moved downward to the bottom of the $^{20}$Ne burning region to the depth at which $^{20}$Ne is completely consumed by nuclear burning. The case for $^{16}$O is similar, but can be seen operating at the lower convective boundary as well. 

{\bf Nuclear burning:} $ 2 \eht{X''_i \rho \dot{X}_i^{\rm nuc}} $. Nuclear burning results in the growth or decay of fluctuations due to nuclear reactions. In the case of $^{20}$Ne, the abundance is depleted through nuclear burning and so this term acts as a sink for the variance that was transported or produced in the $^{20}$Ne burning zone.

{\bf Numerical Residual (dissipation):} $\mathcal{N}_{\sigma i}$. The dashed black
curve in Figure~\ref{fig:sigma-transport} shows the residual
term. The thin solid black curves shows a model of the scalar dissipation which assumes that
the dissipation timescale is equal to the Kolmogorov damping timescale, defined by
\begin{equation}
\tau_L \equiv \left( \frac{\nu}{\epsilon} \right)^{1/2} \equiv \frac{\fht{k}}{\epsilon_k}
\end{equation}
where $\nu$ is viscosity, $\epsilon$ is the turbulent kinetic energy
transfer rate, and $\fht{k} = 1/2\fht{u''_i u''_i}$ is the Favrian specific turbulent kinetic energy mentioned also in Section \ref{sect:convective-reactive-mixing}. Here, $\epsilon_k$ is
the dissipation rate of $\fht{k}$, which can be approximated by $\epsilon_k
\sim u'^3_{rms}/L$, where $L$ is the scale of the flow \ie essentially the size
of convection zone \citep{ArnettMeakin2009, Cristini2017}.

These timescales allow us to test if our simulation results
  agree with Kolmogorov's first similarity hypothesis, which states that
in turbulent flows with high Reynolds numbers, the statistics of the motion
at the dissipation scale has a universal form that is uniquely determined
by $\nu$ and $\epsilon$ \citep{Pope2000}. We do indeed find that the scalar dissipation rates in 
our simulation, calculated using $\tau_L$ as defined above, provides a good fit to our data (Fig.~\ref{fig:sigma-transport}), with
\begin{align}
\mathcal N_{\sigma i} \sim \frac{\eht{\rho} \sigma_i}{\tau_L} \sim \eht{\rho} \sigma_i \frac{\epsilon_k}{\fht{k}} \sim \eht{\rho} \sigma_i \frac{1}{L} \frac{u'^3_{rms}}{\fht{k}}.
\label{eq:nsigma1}
\end{align}
This result supports the physical consistency of the sub-grid scale numerics of our
simulation and the ILES approach, even at rather low spatial resolution and effective Reynolds
numbers, Re$^{{\rm eff}}$. In our simulation,  Re$^{{\rm eff}} \sim N_{c}^{4/3} \sim 1440$, where $N_c \sim 234$ is the
number of radial grid points across the convection zone. That the
  composition variance dissipation rate ${\mathcal N_{\sigma i}}$ is
  directly proportial to $\eht{\rho} \sigma_i / \tau_L$
  (Eq.~\ref{eq:nsigma1}) is consistent with mixing models of molecular
  diffusion. This further confirms the assumption that the rate of scalar
  mixing is determined by scalar variance production at large scales
  \citep{fox1995computational}.

 Despite the inviscid assumption of the Euler equations, scalar
   dissipation in the composition variance will occur because of the
   presence of a numerical mass flux $\mathcal Y$. This will modify all
   continuity equations such that they have a form:
    
\begin{align}
\partial_t \big( \rho X_{i} \big) = \nabla \cdot \big( \rho u_i X_i + {\mathcal Y} \big) + \rho \dot{X}_{i}^{\rm nuc}.
\label{eq:cont3}
\end{align}  
    
\par In these equations $\nabla \cdot \mathcal Y$ is therefore a
conservative redistribution term that will lead to smearing effects,
i.e. numerical diffusion. Mass conservation is however still maintained
across the domain. We use these equations to represent the modified
equations being solved by the numerical algorithm.







\subsection{Effects of Resolution}
\label{sect:resolution}

As mentioned above, the effective Reynolds number of the flow in our simulation is
  Re$^{{\rm eff}} \sim 1440$ and it is thus expected to be
  in the turbulent regime. This is supported by our finding that the
  variance dissipation is proportional to the variance over the large eddy
  timescale -- a phenomenology expected when turbulence is developed and
  has an inertial range.

Our previous work utilising RA-ILES (``RANS'') 
analysis shows that the qualitative
analysis of the {\it interior} of convection zones is not strongly affected
by resolution, for a wide range of effective Reynolds numbers (Re$^{{\rm eff}}
\sim$ 500 to 3300; \citealt{VialletMeakin2013,mmva14,Cristini2017}). This shows that our
RA-ILES technique is useful even at relatively low resolution -- at least for PPM hydrodynamics. 
Moreover, the RA-ILES 
approach can be used to assess resolution dependence and convergence 
by analyzing the residuals defined in by the mean-field equations.

The {\it boundaries} of convection zones, on the other hand, are generally
under-resolved in stellar interior simulations.  This is readily
identifiable -- and quantified -- in the RA-ILES residuals (eg.  $\mathcal{N}$
in Fig.~\ref{fig:composition-transport}, \ref{fig:flux-transport}). In the
current simulation, the lower boundary has a large `spike' in
residuals. This is a turbulent boundary layer, and is relatively thin, with
steep gradients in various quantities. We have previously shown that these
residual spikes diminish with increasing resolution
\citep{Cristini2017}. The top boundary also shows a spike in the residuals,
but is much less pronounced because of the shallower gradients present
there -- it is a `softer' boundary.

While these residuals must affect the quantitative rate of entrainment, 
we do not expect the qualitative features of the convective-reactive flow to be strongly
impacted, especially since the residuals become negligible away from the
convective boundaries.

\section{Summary}
\label{sect:summary}

We have analyzed coupled turbulence and nuclear burning
in a 3D hydrodynamic implicit large eddy simulation (ILES)  of an oxygen burning shell in the core of a $23~\Msun$ supernova progenitor star. 

\subsection{Hydrodynamic simulation}\label{sect:hydrosummary}
 
Our initial stellar structure encompassed a single oxygen-burning
convection zone with a stable Ne-burning layer above.  After an initial
transient episode which led to a gentle merging of the O-burning convective
shell with the Ne-burning shell, a quasi-steady layered-shell state was
formed.  Two layers of stable nuclear burning were set up in the joined
velocity field of a single convection zone, with $^{16}$O dominating the
nuclear energy production at the bottom of the convection zone, but
$^{20}$Ne dominating in the upper layers.

The quasi-static state showed a Ne-burning rate in the convection zone up
to seven orders of magnitude higher than in the initial 1D model. This was
due to a steady entrainment of $^{20}$Ne into the hotter, oxygen burning
shell.

Further, neon burning was found to occur over a wide region in the
convection zone. This is in contrast to the situation found in 1D codes for
similar convective-reactive cases, for example during proton ingestion into
He burning regions in low-mass stars. As a consequence of limitations in
their mixing algorithms, 1D codes generally show burning in a very thin
layer, at a depth where the ratio of the burning timescale equals
that of the convective turnover timescale, i.e. where the Damk\"ohler
number equals one. In our simulation this defines only the top of a wider
burning region, and the peak of the burning is well below this point
(Fig.~\ref{fig:convective-reactive2}).

 \subsection{RA-ILES analysis}

The key results presented in this paper are based on
a Reynolds-Averaged ILES (RA-ILES) analysis of the
composition-related mean fields found for our 3D ILES simulation data, and extend our previous investigations which used the turbulent kinetic energy (TKE) equation.

We develop RA-ILES equations (\S\ref{sect:rans-definitions}) which describe the time evolution of mean
composition, turbulent composition flux, and composition variance for each isotope present in the stellar plasma. 
The equations derived are exact and do not employ approximations to the extent
that the continuum approximation is appropriate and flow features are
resolved. This contrasts with use of closure relationships and truncations
of the RANS equations to construct models of turbulence. {\em The
RANS equations, when closed by 3D ILES numerical simulations (RA-ILES), are able to
represent the full range of hydrodynamical behavior that is present in a
stellar interior simulation. }

Our detailed 
RA-ILES analysis revealed, over the timescale of the simulation ($\sim$4 turnovers):

\begin{itemize}
   \item A simple balance between turbulent transport and burning in the convection
     zone is maintained, whereby any amount of material entrained into the convection
     zone is counterbalanced by nuclear burning (\S\ref{sec:meancomp}, Fig.\ref{fig:composition-transport}).
   \item The turbulent composition flux 
    appears to be quasi-static due to a complex balance between
     turbulent production, nuclear burning, turbulent transport,
     pressure force coupling, and centrifugal forces (\S\ref{sec:compflux}, Fig.\ref{fig:flux-transport}).
   \item Composition variance provides information about composition fluctuation
     amplitudes, and the residual of the composition variance mean field equation then
     gives direct information about numerical dissipation timescales (\S\ref{results:variance}). 
     \end{itemize}


\subsection{Dissipation and ILES}
The residual of the composition variance equation implies a match with the
Kolmogorov damping timescale (\S\ref{results:variance},
Fig.\ref{fig:sigma-transport}). This independently confirms a similar
result found using the turbulent kinetic energy equation
\citep{MeakinArnett2007,ArnettMeakin2010}, and indicates a deep consistency
in our ILES approach. In the simulation, the dissipation of both turbulent
kinetic energy and composition variance occurs at the sub-grid level, as
desired.  This should be true of the entropy and temperature variances as
well, because ILES joins onto the inertial range of the turbulent cascade.


\subsection{Implications for Stellar Evolution Theory}

As discussed above (\S\ref{sect:hydrosummary}) we have identified very
substantial differences between 1D stellar evolution modelling and our 3D
modelling. These findings indicate that if shell mergers and entrainment
do happen in real stars, 1D evolutionary models that do not include these
phenomena are likely to be substantially in error. Moreover, based on our
finding of the way in which ingested fuel burns in the convective zone,
it appears that even if 1D models do incorporate shell mergers and
entrainment, current 1D theory would be unable to model them reliably.

The differences identified here will need to be addressed to improve
modeling of convective-reactive phases. If not addressed the errors from
these events will accumulate through the subsequent evolutionary phases and
will cause uncertainty in, for example, pre-supernova evolution and the 3D
explosion models that rely on the 1D models for their initial states.

Further 3D modeling -- and, just as importantly, detailed analysis such as
that presented in this paper -- is vital to guide these improvements by
illuminating the physical processes at play.

%
%
%
%
%
%
%
%

\section*{Acknowledgements}

MM acknowledges support of the Los Alamos National Laboratory where he was
a postdoctoral research fellow and where much of this research was done.
CAM acknowledges support from Los Alamos National Laboratory as well as the
University of Arizona. SWC acknowledges federal funding from the Australian
Research Council though the Future Fellowship grant entitled ``Where are
the Convective Boundaries in Stars?'' (FT160100046). Part of this work was
supported by computational resources provided by the Australian Government
through NCI via the National Computational Merit Allocation Scheme (project
ew6).  WDA acknowledges support from the Theoretical Astrophysics Program
(TAP) and Steward Observatory, where these simulations and first analyses
were begun. Authors acknowledge Raphael Hirschi for his suggestion of
  the RA-ILES acrononym used to describe our analysis.




\bibliographystyle{mnras}
\bibliography{referenc} 



\appendix

\section{Approximate rates of energy generation}\label{nuclearapprox}

The nuclear burning of carbon, neon, oxygen and silicon during advanced evolutionary stages of massive stars involve increasingly extensive nuclear reaction networks, which have been analyzed and reduced to much simpler equivalent forms \citep{wda96}. These equivalent nuclear reactions (see \citealt{wdaHe72,wdaC72,wdaNe74,wdaO74} for original work),  allow calculation of approximate formulas for nuclear energy production rates coming from each of these equivalent reaction links, and are useful for understanding the energy generation. {\em Our simulations actually use the 25-species network in each zone. Here we give simpler approximations which are useful for analysis; see Fig.~\ref{fig:convective-reactive2}.}

Nuclear energy production due to $^{12}$C burning goes primarily to unstable highly excited state of $^{24}$Mg which decays further through neutron, $\alpha$ and proton channel to Mg$^{23}$, $^{20}$Ne and Na$^{23}$ and its rate (in erg g$^{-1}$ s$^{-1}$) can be approximated by \citep{wdaC72}:
\begin{align}
\dot{\epsilon}_{c12}^{\rm nuc} \approx 4.8 \times 10^{18} \ Y_{c12}^2 \ \rho \ \lambda_{c12,c12} , 
\label{eq:carbon-burning}
\end{align}
where $\lambda_{c12,c12}$ is reaction rate for $\rm ^{12}C(^{12}C,\gamma)^{24}Mg$  \citet{Caughlan1988}.

Neon burns in massive stars primarily by reaction 2 Ne$^{20} \rightarrow$ $^{16}$O + $^{24}$Mg, assuming approximate equilibrium between one photo-disintegrated $^{20}$Ne into $^{16}$O and $^4$He and one $\alpha$ capture on the new oxygen restoring $^{20}$Ne. At this point, one $^4$He starts to add to $^{20}$Ne and produces $^{24}$Mg. The net effect is that for each 2 neon nuclei there is one new nucleus of $^{16}$O and one of $^{24}$Mg produced. Its nuclear production rate (in erg g$^{-1}$ s$^{-1}$) can be approximated by \citep{wdaNe74}:
\begin{align}
\dot{\epsilon}_{ne20}^{\rm nuc} \approx 2.5 \times 10^{29} \ T_9^{(3/2)} \ \frac{Y_{ne20}^2}{Y_{o16}} \ \lambda^{\alpha \gamma}_{ne20} \ e^{-(54.89/T_9)}
\label{eq:neon-burning}
\end{align}
where $\lambda^{\alpha \gamma}_{ne20}$ is the reaction rate for $\rm ^{20}Ne(\alpha,\gamma)^{24}Mg$ \citet{Caughlan1988} and $T_9$ is temperature $T$ divided by 10$^9\rm K$. The equilibrium in the neon burning holds in our network, as indicated by a proximity of the neon nuclear timescales for $\rm Ne^{20} \rightarrow He^4 + O^{16}$ and $\rm He^4 + O^{16} \rightarrow Ne^{20}$ shown in Figure \ref{fig:convective-reactive2}.

Oxygen burning produces compound nuclear states of $^{32}$S which may decay by neutron, proton, deuteron or $\alpha$ channel into $^{31}$S, $^{31}$P, $^{30}$P or $^{28}$Si but the $n$, $p$ and $\alpha$ are recaptured, giving a distribution of nuclei from $\rm ^{28}Si$ to Ca isotopes \citep{ta70,wdaO74}. 
 The energy production rate (in erg g$^{-1}$ s$^{-1}$) can be approximated by:
\begin{align}
\dot{\epsilon}_{o16}^{\rm nuc} \approx 8 \times 10^{18} \ Y_{o16}^2 \ \rho \ \lambda_{o16,o16} 
\label{eq:oxygen-burning}
\end{align}
$\lambda_{o16,o16}$ is reaction rate for $\rm ^{16}O(^{16}O,\gamma)^{32}S$ \citet{Caughlan1988}.

Simplified view of silicon burning involves 2 nuclei of $^{28}$Si producing one $^{56}$Ni nucleus via specific sequence of photo-disintegration and fusion reactions. The critical reaction that allows the photodisintegration of $\rm ^{28}$Si is $\rm ^{24}Mg(\gamma,\alpha)^{20}Ne$. The energy production rate can during such silicon burning be approximated by:
\begin{align}
\dot{\epsilon}_{si28}^{\rm nuc} \approx 1.8 \times 10^{28} \ T_9^{3/2} \ X_{si28} \ \lambda^{\alpha \gamma}_{ne20} \ e^{-142.07/T_9} 
\label{eq:silicon-burning}
\end{align}
\citep{Clayton1983,Woosley2002}.









\bsp	
\label{lastpage}
\end{document}